\documentclass[aps,floats,pre,twocolumn]{revtex4}

\usepackage{amsfonts,amsmath,amssymb}
\usepackage{bm}
\usepackage{dcolumn}
\usepackage{epsfig}
\usepackage{latexsym}
\usepackage{amsthm}
\usepackage{enumerate}

\newcommand{\U}{{\boldsymbol{U}}}
\newcommand{\OM}{{\boldsymbol{\omega}}}
\newcommand{\OT}{{\boldsymbol{\theta}}}
\newcommand{\CS}{{\boldsymbol{{\cal S}}}}
\newcommand{\BU}{{\boldsymbol{{\cal U}}}}
\newcommand{\RH}{{\boldsymbol{\rho}}}
\newcommand{\pphi}{{\pmb\phi}}
\newcommand{\ppsi}{{\pmb\psi}}

\newcommand{\ZM}{{\mathbb Z}}
\newcommand{\RM}{{\mathbb R}}
\newcommand{\CM}{{\mathbb C}}

\newcommand{\TM}{\mathbb T}
\newcommand{\Id}{\mathbb I}

\newcommand{\eig}{{\rm Eig}}

\newtheorem{proposi}{Proposition}

\newtheorem{coro}{Corollary}

\usepackage{soul}
\usepackage[dvipsnames]{xcolor}

\begin{document}

\title{Self-duality triggered dynamical transition}

\author{Italo Guarneri,$^{1}$ Chushun Tian,$^{2}$ and Jiao Wang$^{3}$}

\affiliation{$^1$Center for Nonlinear and Complex Systems, Universit$\grave{a}$ degli Studi dell'Insubria, via Valleggio 11, 22100 Como, Italy\\
$^2$CAS Key Laboratory of Theoretical Physics and Institute
of Theoretical Physics, Chinese Academy of Sciences, Beijing 100190, China\\
$^3$Department of Physics and Key Laboratory of Low Dimensional Condensed Matter Physics (Department of Education of Fujian Province), Xiamen University, Xiamen 361005, Fujian, China}

\begin{abstract}

A basic result about the dynamics of spinless quantum systems is that the Maryland model exhibits dynamical localization in any dimension. Here we implement mathematical spectral theory and numerical experiments to show that this result does not hold, when the 2-dimensional Maryland model is endowed with spin 1/2 -- hereafter dubbed spin-Maryland (SM) model. Instead, in a family of SM models, tuning the (effective) Planck constant drives dynamical localization--delocalization transitions of topological nature. These transitions are triggered by the self-duality, a symmetry generated by some transformation in the parameter -- the inverse Planck constant -- space. This provides significant insights to new dynamical phenomena such as what occur in the spinful quantum kicked rotor.

\end{abstract}

\date{\today}
 -

\maketitle

\section{Introduction}
\label{sec:introduction}
The dynamics of complex quantum systems is currently a very active area of research.  The quantum kicked rotor (QKR) \cite{Casati79,Fishman82a} -- a rotating particle subject to time-periodic ``kicks" --  is  a paradigmatic model, which has played a key role in the development  of this field; moreover, it has allowed for experimental realizations \cite{Fishman10,Raizen11,Garreau16}. A prototypical feature of the  QKR is dynamical localization \cite{Casati79}: though the  classical version of the QKR exhibits unbounded chaotic diffusion in momentum space \cite{Chirikov79},  the QKR  momentum remains bounded in time. Thorough investigations of the analogy \cite{Fishman82a} between  dynamical localization and Anderson localization in disordered solids have given rise to a highly
interdisciplinary field; see Refs.~\cite{Fishman10,Raizen11,Garreau16,Izrailev90,scho} for reviews of different aspects. Several variants of the QKR have been examined, including  2-dimensional ($2$D) versions
\cite{Fishman88} and 1-dimensional ($1$D) versions with spin \cite{Scharf89}. Such variants  did not command major changes of the general picture of dynamical localization. A surprising result has been  instead observed with a recent
$2$D variant endowed with spin $1/2$ \cite{Tian14,Tian16}.  By tuning the
Planck constant $\hbar_e$, a sequence of dynamical transitions arise. At certain critical values of $\hbar_e$ the wavepacket spreads diffusively, leading to dynamical delocalization, while in between different  critical values dynamical localization persists. This phenomenon is remindful of the integer quantum Hall effect (IQHE) \cite{Klitzing80}, with which it shares  the phase diagram and the universality class of phase transition \cite{Khmelnitskii83,Pruisken07}. In particular, dynamical localization corresponds to a quantum Hall insulator, with a topological angle emerging from a supersymmetric field-theoretic description of the spinful QKR, and the dynamical localization--delocalization transition corresponds to the Hall plateau transition. This finding  suggests  that spin is a seed for a wealth of novel dynamical phenomena, which have deep connections to topological quantum matter but go beyond the Thouless-Kohmoto-Nightingale-den Nijs-like topological theory for various driven quantum systems \cite{Lebouf90,Galitski11,Gong12,Gong18,Dana19}.

A fundamental difficulty in assimilating dynamical
localization to Anderson localization is the absence of randomnness/disorder in QKR.  Appeal has been made to a pseudorandom character \cite{Brenner92}
of the eigenphases of free rotation, and/or to a chaotic nature of the early QKR rotor evolution \cite{Izrailev90,Tian04,Tian05}, inherited from the classical limit. However it is now generally acknowledged, that quantum localization is a  general phenomenon, of which Anderson and
dynamical
localization are different occurrences. Important in this sense has been  a modification of the original QKR, the so-called Maryland model  \cite{Fishman82,Fishman84}, where the free Hamiltonian is linear in momentum, and not quadratic, as it is in the QKR  \cite{footnote_SM}.
Due to this difference, pseudo-randomness is replaced by quasi-periodicity, and, moreover, the model is classically integrable \cite{Berry84}. That notwithstanding, quantum localization still survives, motivating taking the model as a basis to study some new problems in condensed matter \cite{Sarma14,Galitskii16}. However this localization is in a more  elementary form.
In particular, it is not a quantum interference effect, so it cannot be described by the scaling theory of Anderson localization and dynamical localization in QKR \cite{Tian04,Tian05,Anderson79,Larkin79,Garreau18}.

In this paper, we present a SM model that, like its spinless ancestor, though bearing no analogies to crystals, is neither pseudo-random nor chaotic and thus is very different from spinful QKR. We show that a family of SM models preserve some aspects of the similarity between IQHE on the one side and 2D, spin-1/2 QKR on the other. Notably, dynamical localization--delocalization transitions appear at half-integer $\hbar_e^{-1}\equiv \alpha/\pi$, indicating a deeper origin of that similarity. We find that such transition arises from a self-duality. That is, SM systems on one side of a critical point, labelled by the parameter $\alpha$, can be one-to-one mapped onto SM systems on the other side via a unitary transformation, so that they are in a ``duality'' relation; at the critical point the system is mapped onto itself, {\it i.e.} it is self-dual, and as such it bears an emergent unitary symmetry. Thus this transition resembles a very recent discovery of topological sound waves in self-dual mechanical systems \cite{Vitelli20}.

Unlike the original Maryland model, the SM model has no classical
limit ($\hbar_e\to 0$); nevertheless both models are equivalent to certain classical dynamical systems,
which belong to the family of so-called skew products \cite{aar}. In particular,  the SM model is equivalent to
a class of skew-product on $\TM^2\times SU(2)$, which has attracted significant mathematical attention \cite{Eliasson98,Eliasson02,Aldecoa15,Karaliolios18}. We also show that the observed transition finds a counterpart in the equivalent classical system: at critical $\alpha$ the deviation of initially close trajectories grows in time $t$ as $\propto\sqrt{t}$ asymptotically, while at noncritical $\alpha$ the growth of the deviation saturates at long time.

The rest of the paper is organized as follows. In Sec.~\ref{sec:model_theoretical_tool} we first define the SM model. Then we show that, for any $\hbar_e$,
this model has an equivalent classical dynamical system, and thus the SM model can be viewed from both a quantum and a classical viewpoint. A summary of main reults follows. In Sec.~\ref{sec:numerical} we present numerical results. In Sec.~\ref{sec:math_results} we develop the spectral theory of SM models and study its dynamical implications, which provide a theoretical basis for the numerical findings. In Sec.~\ref{sec:duality} we show that transport associated to the quantum picture of SM models is fully equivalent to trajectory stability associated to the classical picture of SM models, and derive a result used in Sec.~\ref{sec:math_results}. The technical proofs of the mathematical results claimed in Sec.~\ref{sec:math_results} are shuffled to Sec.~\ref{sec:proof}. We conclude with Sec.~\ref{sec:con}. Some additional information are provided in Appendixes.

\section{SM model and summary of main results}
\label{sec:model_theoretical_tool}

\subsection{Definitions and basic properties}
\label{sec:SM}

We start with a description of the SM model. Consider a spin $1/2$ quantum particle moving on a $2$-torus ${\mathbb T}^2$ ({\it i.e.}, a $2$-rotor), with a constant angular velocity $\OM\equiv(\omega_1,\omega_2)$. The particle's angular coordinates are denoted by $\OT\equiv(\theta_1,\theta_2)$, and their conjugate momentum operators are denoted
${\hbar_e\boldsymbol{N}}\equiv-i\hbar_e(\partial_{\theta_1},\partial_{\theta_2})$. The state vector  of the particle at time $t$ is a spinor with two complex components,
\begin{equation}\label{eq:spinor}
    \pmb{\psi}_t(\OT)=\left(
                  \begin{array}{c}
                    \psi_{1,t}(\OT) \\
                    \psi_{2,t}(\OT) \\
                  \end{array}
                \right),
\end{equation}
which belongs to the Hilbert space ${\cal H}\equiv L^2({\mathbb T}^2)\otimes\CM^2$. The particle is kicked periodically in time; with the period set to unity, at all integer times $t\in \mathbb{Z}$
it undergoes instantaneous pulses, or kicks, that prompt sudden jumps of its momentum and simultaneously rotate its spin in a way, that depends on the angular coordinates. Such kicks are produced by switching on  an external ``potential" $\pmb V(\OT)=\sum_{k=1}^3V_k(\OT)\pmb\sigma_k$. Here $\pmb\sigma_k$ are the Pauli matrices and the $V_k$ are some functions $\mathbb{T}^2\rightarrow \mathbb{R}$.  The quantum dynamics follows the Schr\"odinger equation:
\begin{eqnarray}
i\hbar_e \partial_t \pmb{\psi}_t(\OT) = \boldsymbol{H}(t) \pmb{\psi}_t(\OT)\,,
\qquad\qquad\qquad \label{eq:Hamiltonian}\\
\boldsymbol{H}(t) = \hbar_e\OM\pmb \cdot \boldsymbol{N} + V(\OT)\sum_{s\in \mathbb{Z}}\delta(t-s).\nonumber
\end{eqnarray}
 The first term of the Hamiltonian $\boldsymbol{H}(t)$ corresponds to rotation over the torus $\mathbb{T}^2$
 with constant angular velocity and the second to the kick. The dynamics  from immediately after a kick to immediately after the next is described by:
\begin{equation}
\label{eq:qhe}
 \forall t\in \mathbb{Z}:\,\pmb{\psi}_{t+1}(\OT)\;=\;\U_{\alpha,\OM} \pmb{\psi}_{t}(\OT),\quad \U_{\alpha,\OM}:= \boldsymbol{M}_\alpha\boldsymbol{T}_{\OM},
\end{equation}
which defines the Floquet operator $\U_{\alpha,\OM}$. Here $\boldsymbol{T}_{\OM}$ is the translation operator: $\boldsymbol{T}_{\OM}\pmb{\psi}(\OT)\equiv \pmb{\psi}(\pmb\tau_{\pmb\omega}^{-1}\OT)$ where $\pmb\tau_{\pmb\omega}:\OT
\mapsto
\pmb \OT+\pmb\omega$ is translation by $\pmb\omega$ along the torus $\TM^2$. Instead $\boldsymbol{M}_\alpha=e^{-i\frac{\alpha}{\pi}\pmb V(\OT)}$ is a smooth map ${\mathbb T}^2\to SU(2)$;
so $\pmb V(\OT)$ is a smooth map from $\TM^2$ to the $2\times 2$ self-adjoint matrices with null trace. It will be assumed to have  a constant Hilbert-Schmidt norm such that ${\rm tr}(\pmb V(\OT)^2)=2\pi^2$ , so it can be written in the general form:
\begin{equation}\label{eq:spec_V}
  \pmb V(\OT)=\pi\sum_{k=1}^3d_k(\OT)\pmb\sigma_k\;,
\end{equation}
which with some abuse of notation will be concisely written in the form  $\pi\boldsymbol{d}(\OT)\pmb{\cdot}\boldsymbol{\sigma}$. Here
${\pmb d}(\OT)=(d_1(\OT),d_2(\OT),d_3(\OT))$ is a fixed map from $\TM^2$ to the unit real $2$-sphere $\mathbb{S}^2$.
 Equation (\ref{eq:Hamiltonian}) or its discrete-time version Eq.~(\ref{eq:qhe}) define our general SM model.

The following basic property will be extensively used in the following:
\begin{equation}
\label{eq:rholem}
\RH\boldsymbol{M}_\alpha\RH^\dagger\;=\;\boldsymbol{M}_\alpha,\quad \RH\U_{\alpha,\OM}\RH^\dagger
\;=\;\U_{\alpha,\OM}\;
\end{equation}
where $\RH$ is the standard time-reversal operator for spin-$1/2$ particles; $\RH := -i\pmb\sigma_2 C$ (where $C$  denotes complex conjugation in the coordinate representation). It is the  antiunitary map that is defined in $\cal H$ by
\begin{eqnarray}
\label{eq:rho}
  \RH: \left(\begin{array}{c}
          \psi_1 \\
          \psi_2
        \end{array}\right)
  \mapsto \left(\begin{array}{c}
          -\overline{\psi_2} \\
          \overline{\psi_1}
        \end{array}\right)
\end{eqnarray}
The ordinary Maryland model \cite{Fishman82,Fishman84} is recovered from the present one, whenever  the potential $\boldsymbol{V}(\OT)$ is invariant under  spin rotations, {\it i.e.}, $\boldsymbol{V}(\OT)$ does not couple the angular and  spin degrees of freedom. As shown below, thanks to this difference the SM model exhibits a much more intriguing phenomenology than the ordinary Maryland model.

On the other hand, the SM model crucially differs from the spinful QKR \cite{Tian14,Tian16} in that its free Hamiltonian between kicks is linear in $n_1$, while for the spinful QKR the free Hamiltonian is quadratic in $n_1$. It is  well known that  kicked systems with a quadratic free Hamiltonian may exhibit a chaotic behavior (in some sense), which cannot be observed with a linear one.

\subsection{Equivalent classical system}
\label{sec:SM_classical}

Thanks to the linear momentum dependence of the free Hamiltonian, the SM model Eq.~(\ref{eq:qhe}), though of quantum origin, can be translated into a classical dynamical system, with the usual technical meaning, {\it i.e.} a discrete time dynamics which is generated by iterating a measure preserving transformation in some phase space. Though mathematically trivial, this quantum-classical juxtaposition illustrates that in SM models  wavepacket propagation
and motion along phase-space trajectories are equivalent pictures of the same dynamics. The phase space  is  $\Omega\equiv {\mathbb T}^2\times\CM_1^2$, where $\CM_1^2$ denotes the unit sphere in $\mathbb{C}^2$, and  the map is defined as follows:
 \begin{gather}
 \CS_{\alpha,\OM}: (\OT,\pmb{\phi})\mapsto(\boldsymbol{\tau}_{\pmb\omega}\OT,\boldsymbol{M}_\alpha(\boldsymbol{\tau}_{\pmb\omega}\OT)\pmb{\phi})\,
 .
 \label{eq:sysdin}
 \end{gather}
This map  preserves the natural measure $d\mu$ that is defined in $\Omega$ by the product of the Lebesgue measure  $dm$ on $\TM^2$, normalized to unity, and the uniform measure on $\CM^2_1$ [or equivalently the Haar measure on $SU(2)$]. In mathematical terms  the map described by Eq.~(\ref{eq:sysdin}) defines a skew product action of $SU(2)$ \cite{Eliasson98,Eliasson02,Aldecoa15,Karaliolios18}. The corresponding Perron-Frobenius operator, denoted by $\BU_{\alpha,\OM}$,
acts on functions $F\in {\mathfrak H}\equiv L^2(\Omega,d\mu)$ as follows:
\begin{eqnarray}
\label{eq:FP}
  \BU_{\alpha,\OM}F(\OT,\pmb{\phi}) &=& F(\CS_{\alpha,\OM}^{-1}(\OT,\pmb{\phi}))\nonumber\\
  &=& F(\boldsymbol{\tau}^{-1}_{\pmb\omega}\OT\;,\;
 \pmb M_\alpha(\OT)^{-1}\pmb{\phi})\;.
\end{eqnarray}
It preserves the scalar product in $\mathfrak H$, {\it i.e.} $\forall F',F''\in \mathfrak H$,
\begin{eqnarray}
\label{eq:product}
  \langle F'|F''\rangle_{\mathfrak  H} &:=& \int_{\Omega}d\mu(\OT,\pmb{\phi})\;\overline{F'(\OT,\pmb{\phi})}F''(\OT,\pmb{\phi})\nonumber\\
  &=& \langle{\BU_{\alpha,\OM} F'}|{\BU_{\alpha,\OM} F''}\rangle_{\mathfrak H}\;.
\end{eqnarray}
The actual link between the ``quantum'' SM model and its ``classical'' equivalent is as follows. Given a $\pmb{\psi}\in\cal H$, let
\begin{equation}\label{eq:def_F}
  F_{\pmb{\psi}}(\OT,\pmb{\phi}):=\pmb{\psi}(\OT)\pmb \cdot\pmb{\phi}.
\end{equation}
Here the dot denotes the canonical scalar product of vectors in $\CM^2$: $\pmb{\psi}\pmb \cdot\pmb{\phi}:=\overline{\psi_1}\phi_1+\overline{\psi_2}\phi_2$. Then
 \begin{eqnarray}
 \BU_{\alpha,\OM} F_{\pmb{\psi}}(\OT,\pmb{\phi})\;=\;\U_{\alpha,\OM}\pmb{\psi}(\OT)\pmb \cdot\pmb{\phi}\;=\;F_{\pmb{\psi}'}(\OT,\pmb{\phi})\;,
 \label{eq:conn}
 \end{eqnarray}
where $\pmb{\psi}'=\U_{\alpha,\OM}\,\pmb{\psi}$. This equation formalizes the quantum-classical juxtaposition within the same model.
It is important to emphasize that the quantum system (\ref{eq:qhe}) and the classical dynamical system (\ref{eq:sysdin}) are fully equivalent, and the latter is not related to any notion of classical limit whatsoever: the subscript $\alpha$ -- {\it i.e.} the inverse Planck constant -- in $\CS_{\alpha,\OM}$ is a bookkeeping of this fact. The existence of a classical equivalent of the SM model implies that quantum interference does not play any role in its phenomenology. As mentioned in the introduction, the same is true for the original Maryland model, where the observed localization is not related to the Anderson-like dynamical localization that typically occurs in quantum systems exhibiting chaotic diffusion in the classical limit and results from quantum interference.

The spectral and dynamical properties of the SM models crucially depend on the arithmetic type of the triple
$(\omega_1,\omega_2,\pi)$. In particular, the following well-known fact is of capital importance to the present work: \noindent{\it the $\OM$-shift $\boldsymbol{\tau}_{\pmb\omega}$ is ergodic in ${\mathbb T}^2$ whenever ($\omega_1,\omega_2,\pi$) are an incommensurate triple of real numbers.}
 More detailed arithmetic properties of the triple play important spectral dynamical roles; a few more  comments on this theme are deferred to Sec.~\ref{sec:spec_dyn}. However, for the purposes of the present paper sheer incommensuration will be sufficient. From now on, we shall always consider  incommensurate triples, unless explicitly mentioned otherwise.

\subsection{Outline of main results}
\label{sec:summary}

In this work we implement the quantum-classical juxtaposition to investigate  both mathematically and numerically the dynamical properties of SM. In numerical simulations, we used two different incommensurate frequency vectors $\pmb\omega$, which are specified in Sec.~\ref{sec:trans}.

Upon expanding  $\pmb\psi_t(\OT)$ in the
$2$D Fourier basis  we move  from  the $\OT$-coordinate  representation  to the momentum $\boldsymbol{N}$-representation, where the SM dynamics can be pictured as propagation of a quantum wavepacket over
the $2$D integer lattice $\mathbb{Z}^2$. This defines quantum transport for the SM models in the same way as for the QKR \cite{Casati79} and for the ordinary Maryland model \cite{Fishman82,Fishman84}. A basic probe of the quantum transport associated with the evolution of a state $\ppsi_t$ from an initial state $\ppsi_0$ is
\begin{eqnarray}
\label{eq:moment}
  &&E_{j}(\ppsi_t):=\langle \ppsi_t|n_j^2|\ppsi_t\rangle=\|\partial_{\theta_j}\pmb{\psi}_t\|^2_{\cal H},\,j=1,2 \nonumber\\
  &&E(\ppsi_t)\;=\;E_1(\ppsi_t)\;+\;E_2(\ppsi_t),\;
\end{eqnarray}
which characterizes how the wavepacket spreads in the course of time. Like QKR systems \cite{Casati79,Fishman82a,Tian14,Tian16},
this probe provides an important link between SM models and condensed matter systems. Notably, when $E(\ppsi_t)$ remains bounded as $t\to\infty$, the wavepacket does not spread at long time and remains localized in a finite region in momentum space; no quantum transport arises and this simulates a quantum insulator in condensed matter physics. When either or both $E_{1,2}(\ppsi_t)$ display unbounded growth, the wavepacket propagation is delocalized; and if the asymptotic growth is linear in time, then  quantum diffusive transport arises, and this simulates a quantum normal metal in condensed matter physics. Such different transport properties are mirrored by the spectral properties of the Floquet operators $\U_{\alpha,\OM}$; a detailed analysis is given in Sec.~\ref{sec:math_results}.

We will show that quantum transport in the SM models finds a ``classical" counterpart in the stability of trajectories, {\it i.e.} in the way  initially close trajectories separate in time. This is a fundamental token in the theory of classical dynamical systems,  and a commonly accepted definition of "chaos" applies whenever the divergence of trajectories is exponentially fast. Specifically, the quantum transport in momentum space, probed by $E_{1,2}(\ppsi_t)$ defined by Eq.~(\ref{eq:moment}), has an equivalent description in terms of the instability of trajectories $(\OT_t,\pmb{\phi}_t):=\CS_{\alpha,\OM}^t(\OT,\pmb{\phi})$ in the phase space $\Omega$, in the following sense.
The (linear) stability of a trajectory $(\OT_t,\pmb{\phi}_t)$ is determined by the behavior in time of the derivatives $\partial_{\theta_j}\pmb{\phi}_t(\OT,\pmb{\phi})$ with $j=1,2$; see Sec.~\ref{sec:duality}, where  we prove that if the initial spinor $\ppsi_0(\OT)$ is localized at the origin of momentum space  then the quantum transport which is defined by Eq.~(\ref{eq:moment}) is related to trajectory stability by:
\begin{eqnarray}
\label{eq:Ejt}
  E_j(\ppsi_{-t})
&=&\int_{\Omega}d\mu(\OT,\pmb{\phi})\;\pmb{\vert}\partial_{\theta_j}\pmb{\phi}_t(\OT,\pmb{\phi})\pmb{\vert}^2.
\end{eqnarray}
Spectral implications of this result  are presented  in Sec.~\ref{sec:spec_dyn}.

In  this paper we study three different SM models. Their Floquet operators $\U_{\alpha,\OM}$ share the general form described in Sec.~\ref{sec:SM}. Differences arise from the explicit form of $\boldsymbol{d}(\OT)$:
\begin{itemize}
  \item SM Model I:
\begin{eqnarray}
\label{eq:sm1}
\boldsymbol{d}(\OT)=p(\OT)(\sin(\theta_1),\sin(\theta_2),\beta (1-\cos(\theta_1)-\cos(\theta_2))),
\end{eqnarray}
where $p(\OT)$ is the normalization factor, chosen  such that $\boldsymbol{d}(\OT)$ is a unit vector, and $\beta>0$ is a real parameter.
  \item SM Model II:
\begin{eqnarray}
\label{eq:sm2}
\boldsymbol{d}(\OT)=p(\OT)(\sin(\theta_1),\sin(\theta_2),\mbox{\rm const.}\neq 0);
\end{eqnarray}
\item SM Model III:
\begin{eqnarray}
\label{eq:sm3}
\begin{array}{c}
\boldsymbol{d}(\OT)=p(\OT)(\sin(\theta_1),\sin(\theta_2),\cos(\theta_1)+\cos(\theta_2))\\
{\rm or}\, \boldsymbol{d}(\OT)=p(\OT)(\sin(\theta_1),\sin(\theta_2),0).
\end{array}
\end{eqnarray}
\end{itemize}
An important part of our study is how the transport properties of such models depend on the parameter $\alpha$. In this respect, the following properties:
\begin{equation}\label{eq:period_U}
  \U_{\alpha+2\pi,\OM}=\U_{\alpha,\OM}\,,\,\,\,\,\U_{\alpha+\pi,\OM}=\,-\,\U_{\alpha,\OM}
\end{equation}
which directly follow from the  definition of $\U_{\alpha,\OM}$, hold true for all SM models; so, analysis of SM Models I, II, III can be restricted to $\alpha\in[0,\pi]$. The three models share certain symmetries, which are described in Sec.~\ref{sec:spec_SM}. A fundamental fact is that such symmetries in physical space give rise
to a symmetry in parameter space; notably,  quantum SM systems whose $\alpha\in[0,\pi]$ are symmetric with respect to  $\pi/2$ are in a ``duality"  relation  \cite{nature}. This duality has strong spectral implications, as shown in Sec.~\ref{sec:spec_SM}. Besides, at $\alpha=\pi/2$ it forces the system to be invariant under a unitary symmetry transformation and thus to be self-dual.

The dynamical behaviors of the three SM models are very different:
\begin{itemize}
  \item SM Model I:\\
  We find that for $\alpha\in[0,\pi]$, on the left or on the right of  the critical point $\alpha=\pi/2$,   both $E_{1,2}(\ppsi_t)
  $ remain bounded in time, so  the system is dynamically localized, and localization is exponential in momentum space.  At the critical point $E_{1,2}(\ppsi_t)
  \propto t$ approximately, so the system is dynamically delocalized. In the classical system, the deviation $\delta_t$ of initially close trajectories grows as $\sqrt{t}$ (approximately) at half-integer $\alpha/\pi$,  whereas $\delta_t$ saturates at long time away from the critical point. The transition in quantum transport is thus mirrored by a change of stability of classical trajectories. Interestingly,  sublinear divergence of trajectories is what one should  expect if the classical SM model were known to be ergodic; see Sec.~\ref{sec:duality}. We explain delocalization at the critical value, proving that self-duality is incompatible with dynamical localization. As we will show in Sec.\ref{sec:duality}, this dynamical transition mirrors the transition in the spectral structure of Floquet operator $\U_{\alpha,\OM}$.

  Moreover, it turns out that the dynamically localized phases which are separated by the critical points can be obtained from topologically distinct phases which are found in the spinful QKR \cite{Tian14,Tian16} by deforming that QKR continuously into the considered SM.
  \item SM Model II:\\
  As $\alpha$ varies, dynamical localization is always observed.
  \item SM Model III:\\
  As $\alpha$ varies, dynamical delocalization
  is always observed except at integer $\alpha/\pi$,  where dynamical localization trivially occurs because  $\boldsymbol{M}_\alpha=\pm \mathbb{I}$ there.
\end{itemize}

We see that, due to the presence of spin, SM models exhibit rich dynamical phenomena. This is in sharp contrast to the ordinary Maryland model for which only dynamical localization can occur. In addition,  the Maryland model  inherits dynamical localization from the integrability of the classical dynamics in the limit $\hbar_e\to 0$ \cite{Berry84}; this mechanism for dynamical localization does not apply here, because the SM model is ill-defined in that limit.

\section{Numerical experiments}
\label{sec:numerical}

In this section we describe  numerical results about SM Models I, II and III.

\subsection{$1$D equivalents}
\label{sec:1D}

Direct numerical simulation of the $2$D quantum evolution  requires a $2$D Fourier basis. Thus it cannot be pushed too far in time. This difficulty is greatly softened by a trick which was first used in Ref.~\cite{Casati89}. It consists in replacing the $2$D model described by Eq.~(\ref{eq:qhe}) by two  $1$D models, which  allow  to separately compute $E_{1,2}(\ppsi_t)$ [abbreviated as $E_{1,2}(t)$ throughout this section].  This possibility crucially rests on the $\OM$-shift $\boldsymbol{\tau}_\OM$ in $\mathbb{T}^2$, namely, the first term of $\boldsymbol{H}(t)$ given by Eq.~(\ref{eq:Hamiltonian}), as we will show below.

By performing the transformation:
\begin{eqnarray}\label{eq:2Dto1Dn1}
  \U_{\alpha,\OM}&\rightarrow& e^{t\omega_2\partial_{\theta_2}}\U_{\alpha,\OM}e^{-(t-1)\omega_2\partial_{\theta_2}}\nonumber\\
  &=&\boldsymbol{M}_\alpha(\theta_1,\theta_2+\omega_2 t)e^{-\omega_1\partial_{\theta_1}}=:\tilde{\U}_t,\nonumber\\
  \pmb{\psi}_t&\rightarrow& e^{t\omega_2\partial_{\theta_2}}\pmb{\psi}_t=:\tilde{\pmb{\psi}}_t,
\end{eqnarray}
we rewrite Eq.~(\ref{eq:qhe}) as
\begin{equation}
\label{eq:qhe1Dn1}
 \tilde{\pmb{\psi}}_{t}\;=\;_{\rm T}\!\prod_{s=1}^{t}\tilde{\U}_{s} \tilde{\pmb{\psi}}_{0},
\end{equation}
where $_{\rm T}\!\prod$ denotes a time-ordered product, with the index $s$ increasing from right to left. In Eqs.~(\ref{eq:2Dto1Dn1}) and (\ref{eq:qhe1Dn1}), $\theta_2$ is no longer a dynamical variable,  rather it is an angular parameter. Thus the quantum evolution (\ref{eq:qhe1Dn1}) is $1$D. When the $2$D model is traded for such $1$D model, we have
\begin{equation}\label{eq:moment1}
    E_{1}(t)= -\langle\langle\tilde{\pmb{\psi}}_t|\partial_{\theta_1}^2\tilde{\pmb{\psi}}_t\rangle_{\tilde{\cal H}}\rangle_{\theta_2},
\end{equation}
where the Hilbert space $\tilde{{\cal H}}\equiv L^2({\mathbb T}^1)\otimes\CM^2$ and $\langle\cdot\rangle_{\theta_2}$ denotes the average over $\theta_2$.
The whole derivation is fully symmetric with respect to the indices $1,2$,  so similar equations  are obtained for
$E_2(t)$ by just interchanging indices in the above equations.  Such equations will share the same reference numbers.

Throughout this paper in both $1$D and $2$D simulations we choose initial states which have only the spin-up ($s=1$) component and are zero momentum state.

\subsection{Dynamics of SM Model I}
\label{sec:nummodel1}

\subsubsection{Quantum transport}
\label{sec:trans}

The $1$D quantum evolutions which are described by Eqs.~(\ref{eq:qhe1Dn1}) were numerically simulated up to $t= 10^8$, using $1$D Fast Fourier Transform (FFT) with a basis size $2^{16}=65536$.
The parameter $\beta$ in Eq.~(\ref{eq:sm1}) was set to $0.8$. Two different incommensurate frequency vectors were used: $\OM=(1,2\pi/\sqrt{5})$ and $\OM= (2\pi x, 2\pi x^2)$, where $x$ is the real root
of the equation  $x^3-x-1=0$ \cite{footnote_x}.

\begin{figure}[t]
\vskip-.2cm\hskip-.5cm
\includegraphics[width=\columnwidth]{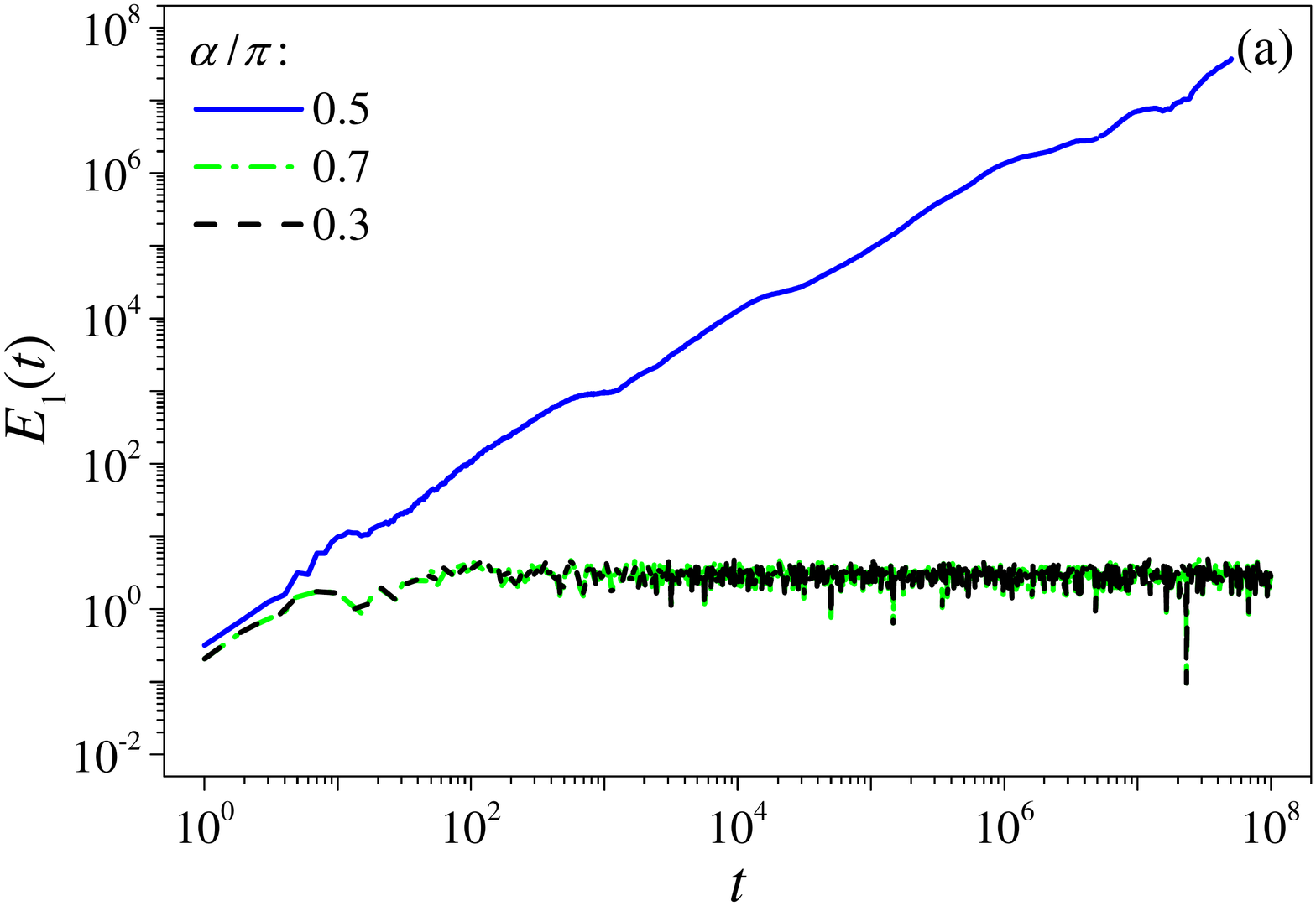}
\vskip-.3cm\hskip-.5cm
\includegraphics[width=\columnwidth]{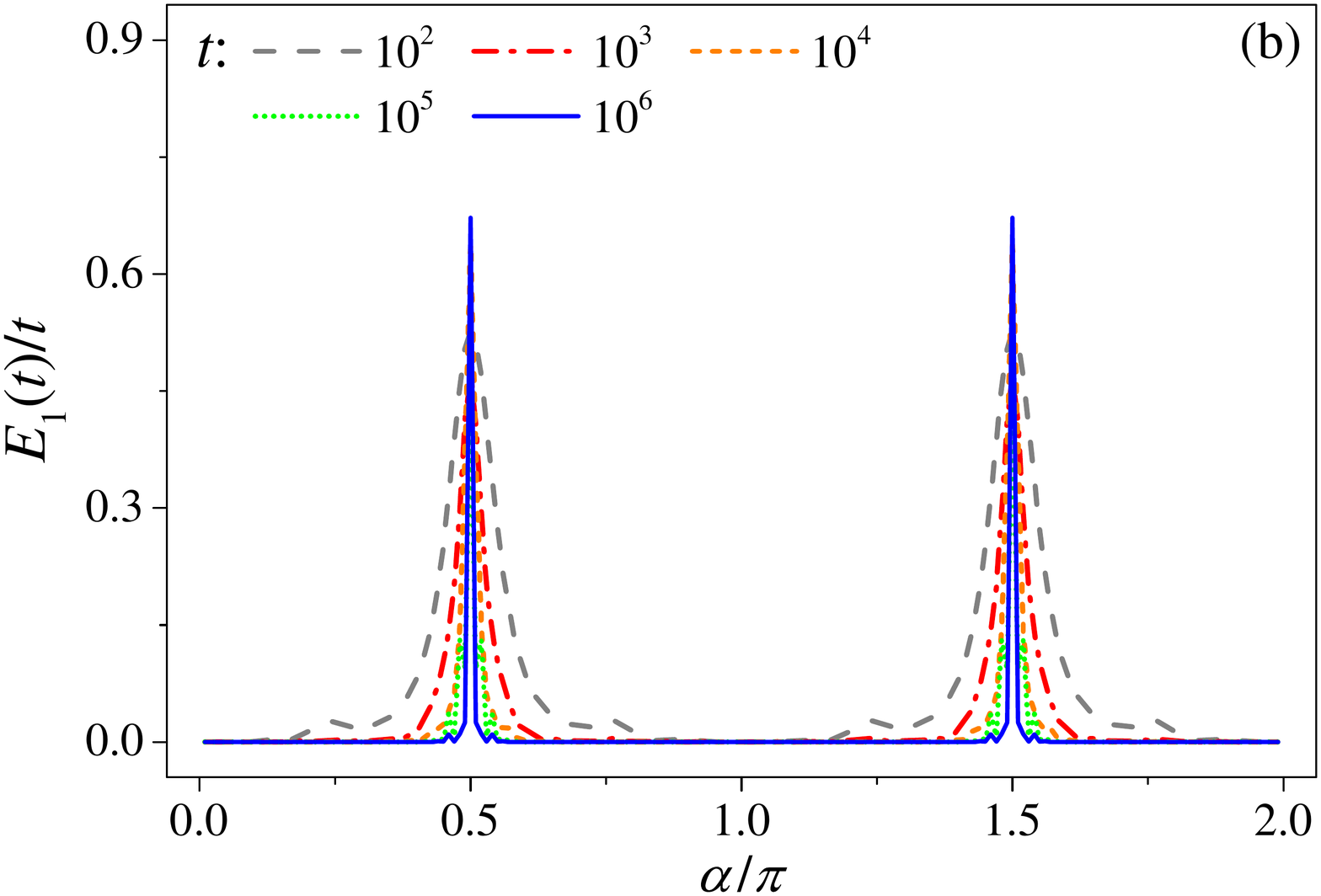}
\vskip-.5cm
\caption{Numerical experiments on SM Model I indicate a dynamical localization--delocalization transition at half-integer $\alpha/\pi$. (a)  shows that $E_1(t)$ grows diffusively at half-integer $\alpha/\pi$ while it saturates at long times away from these points. (b) suggests  a sharp transition at infinite time: the diffusion coefficient is finite at half-integer $\alpha/\pi$ while it vanishes away from the critical points. $\OM=(1,2\pi/\sqrt{5})$.}
\label{fig:2}
\end{figure}

\begin{figure*}[t]
\vskip-.2cm\hskip-.5cm
\includegraphics[width=2\columnwidth]{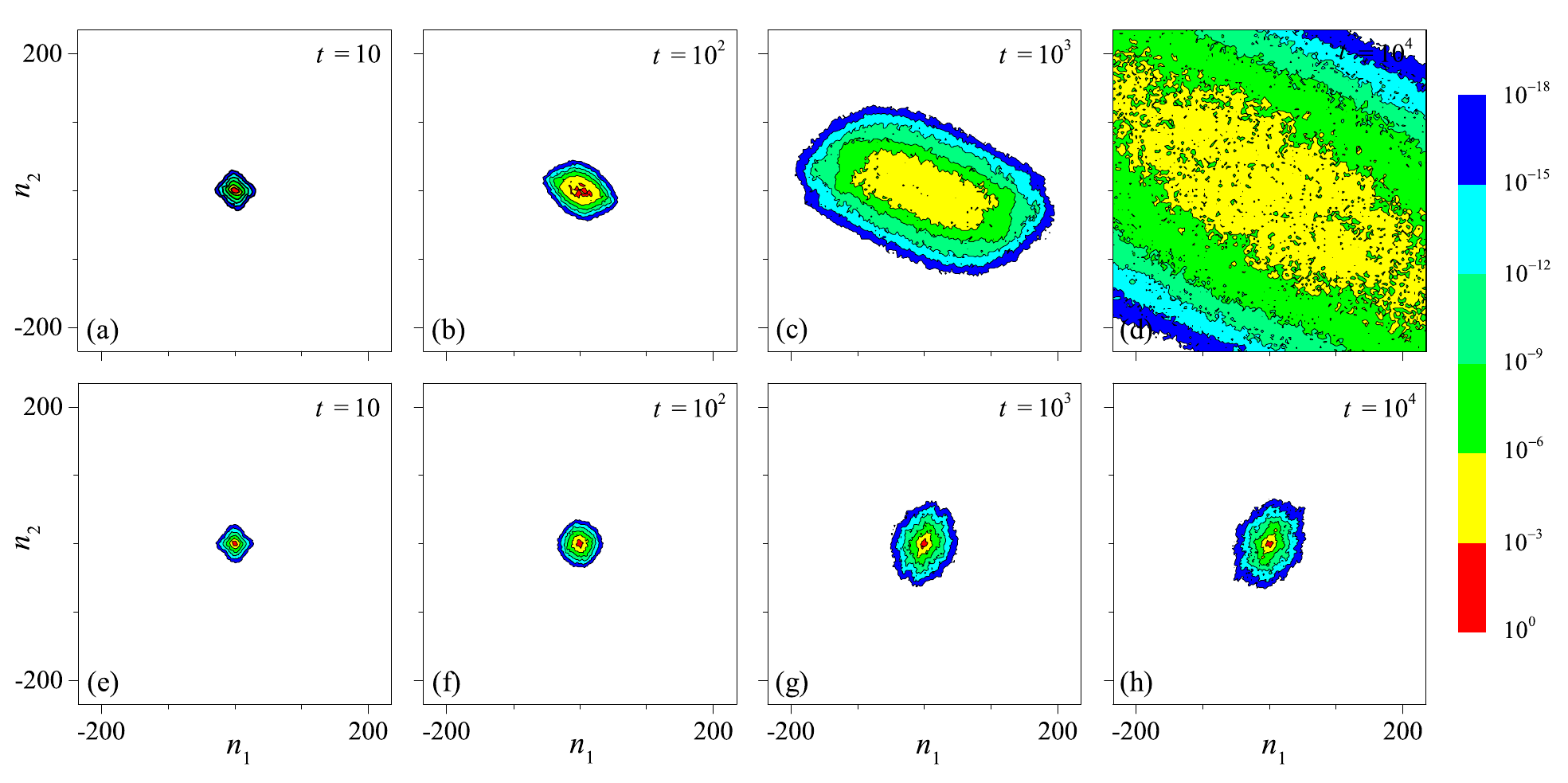}
\caption{A plot of $I_t(n_1,n_2)$ obtained from numerical experiments on SM Model I shows that within the inspected time range, the wavepacket propagation is extended in  momentum space for critical $\alpha/\pi=0.5$ (a-d), while it is restricted to a finite domain for noncritical $\alpha/\pi=0.7$ (e-h). Moreover, the propagation is anisotropic in both cases. $\OM=(1,2\pi/\sqrt{5})$.}
\label{fig:5}
\end{figure*}
Results for $E_1(t)$ and $E_1(t)/t$ are shown in Fig.~\ref{fig:2}. Results for $E_2(t)$ are qualitatively similar. In (a) the cases with $\alpha/\pi=0.3$ and $=0.7$ provide evidence of dynamical localization, and are representative of what was observed away from $\alpha/\pi=0.5$. At $\alpha/\pi=0.5$ or $=1.5$ delocalization was observed, with $E_{1,2}(t)$ increasing approximately
linearly in time.
The $\alpha$-dependence of $E_1(t)/t$ at fixed time is shown in (b).
The appearance of two sharp peaks indicates that delocalization is confined to a narrow vicinity
of $\alpha/\pi=0.5$ or $=1.5$ .  Comparison of plots of $E_1(t)/t$ vs $\alpha/\pi$ at fixed $t$, computed  for  different values of $t$, shows  that as $t$ increases the two peaks become sharper and sharper, suggesting that
they would eventually shrink to two critical points: $\alpha/\pi=1/2$ and $3/2$. This indicates a dynamical localization--delocalization transition at half-integer $\alpha/\pi$. Note that these critical values of $\alpha$ are independent of the choice of $\OM$, so long as $(\OM,\pi)$ is an incommensurate triple. Moreover, the height of the two peaks remains  finite, so it yields  a finite diffusion coefficient $E_1(t)/t={\cal O}(1)$.
The smallness of this coefficient reflects the quantum nature of critical diffusion.

\begin{figure}[h]
\vskip-.2cm\hskip-.5cm
\includegraphics[width=\columnwidth]{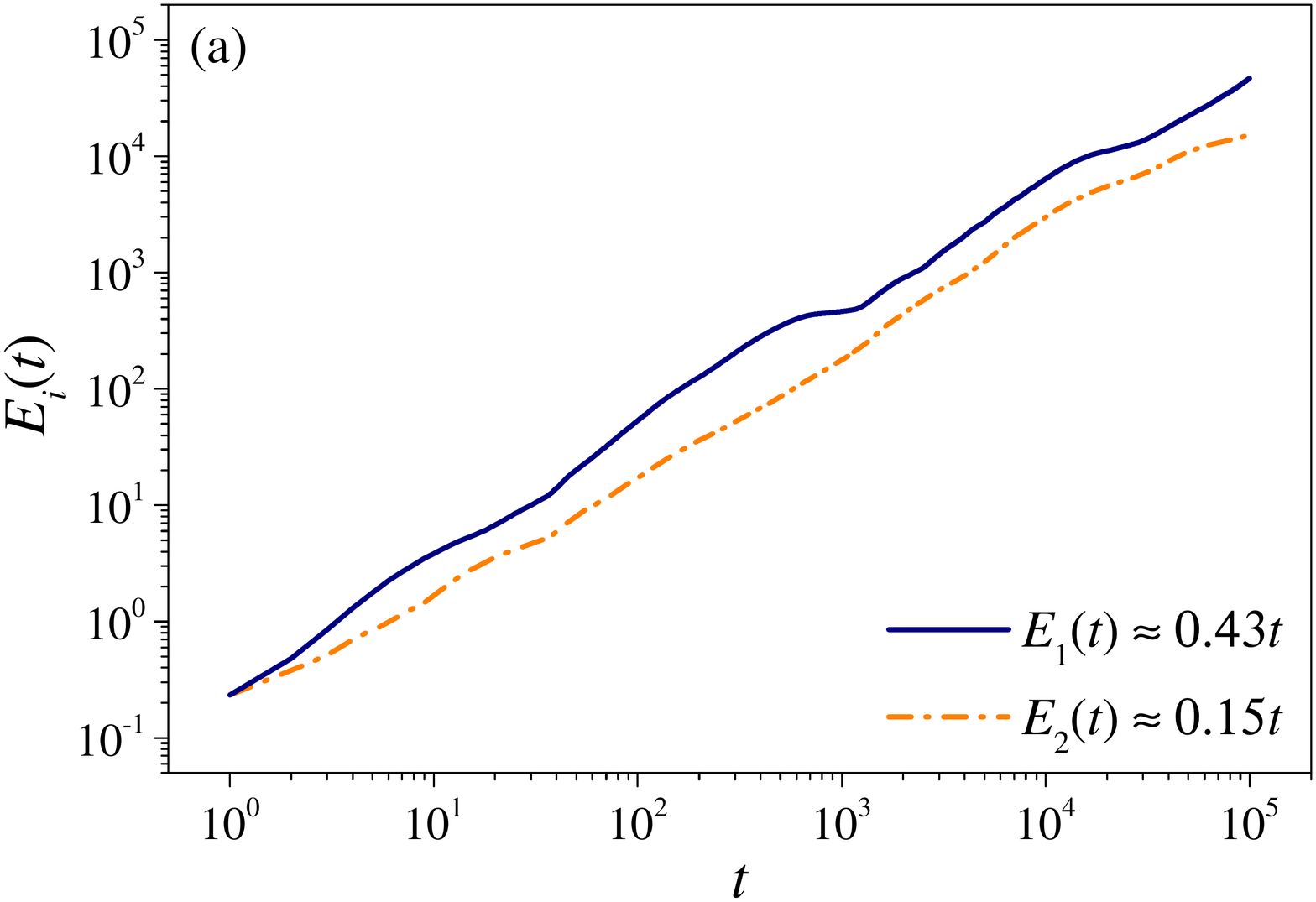}
\vskip-.2cm\hskip-.5cm
\includegraphics[width=\columnwidth]{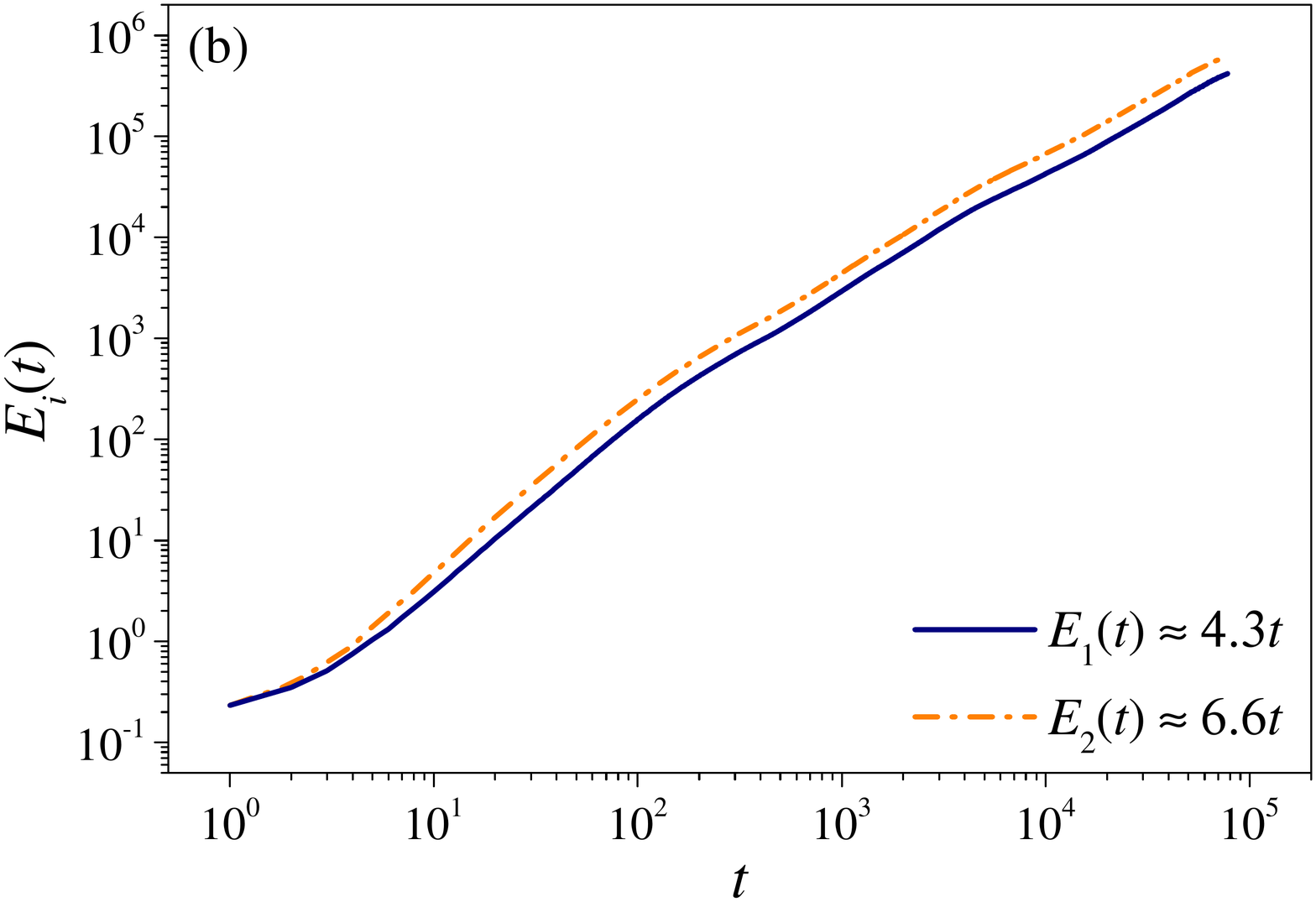}
\caption{Numerical experiments on SM Model I for different frequency vectors $\OM=(1,2\pi/\sqrt{5})$ (a) and $\OM= (2\pi x, 2\pi x^2)$ (see text for the value of $x$) (b) show that the critical diffusion is anisotropic and the diffusion coefficients depend on $\OM$.}
\label{fig:4}
\end{figure}

Results of numerical simulations of the  full $2$D dynamics, using $2$D FFT, are shown in Fig.~\ref{fig:5}  by contour plots of the quantity $I_t(n_1,n_2)\equiv\hat\ppsi_t(n_1,n_2)\pmb\cdot \hat\ppsi_t(n_1,n_2)$,
with $\hat\ppsi_t(n_1,n_2)$ the $2$D Fourier coefficients of $\ppsi_t(\OT)$ . In both the delocalized (a-d) and the localized (e-h) case the wavepacket is seen to propagate in all directions in the momentum space. However, neither diffusion nor localization are isotropic over the inspected range of times. As the quantity $I_t$ is plotted in a logarithmic scale, the approximate equidistance of the level curves in (h) implies exponential localization.
Anisotropy of diffusion is confirmed over much longer times by the plots in Fig.~\ref{fig:4},  which show both $E_1$ and $E_2$ vs time for two different choices of $\pmb\omega$. Diffusion in the $n_1$ and in the $n_2$ direction follows different rates, which  in turn depend on $\pmb\omega$.

To explore the nature of the dynamical localization--delocalization transition, we further carry out numerical experiments on an auxiliary $1$D quantum dynamical system defined as follows:
\begin{gather}
\label{eq:aux1Dn1}
 \check{{\pmb{\psi}}}_{t}\;=\;_{\rm T}\!\prod_{s=1}^{t}\check{{\U}}_{s} \check{\pmb{\psi}}_{0},\nonumber\\
 \check{{\U}}_{t}:=e^{-i\alpha\check{p}(\theta_1,\theta_2+\omega_2t)
 \boldsymbol{d}(\theta_1,\theta_2+\omega_2t)\pmb{\cdot}\boldsymbol{\pmb\sigma}}e^{-i\alpha(h_en_1)^\gamma}
\end{gather}
with $\omega_2$ the second component of the frequency vector $\OM$ in SM models.
Here $\boldsymbol{d}(\pmb{\theta})$ is the same as for SM Model I with $\beta=0.8$, and $\check{p}(\pmb{\theta})=\frac{2}{\pi}\arctan\frac{K}{p(\pmb{\theta})}$ with $K$ a strictly positive parameter and $p(\pmb{\theta})$ given by Eq.~(\ref{eq:sm1}). In addition to $K$, this auxiliary model has  one more strictly positive  parameter $\gamma$. The $1$D equivalent (\ref{eq:qhe1Dn1}) of SM Model I adopted in the above numerical experiments corresponds to $K=+\infty, \gamma=1$, while a special spinful QKR \cite{Beenakker11} that can exhibit a $\hbar_e$-driven IQHE \cite{Tian14,Tian16} corresponds to finite $K$ and $\gamma=2$.
Like the SM models, the quantum dynamics of this system can be probed by $E_{1}(t)= -\langle\langle\check{\pmb{\psi}}_t|\partial_{\theta_1}^2\check{\pmb{\psi}}_t\rangle_{\tilde{\cal H}}\rangle_{\theta_2}$.

\begin{figure}[t]
\vskip-.2cm\hskip-.5cm
\includegraphics[width=\columnwidth]{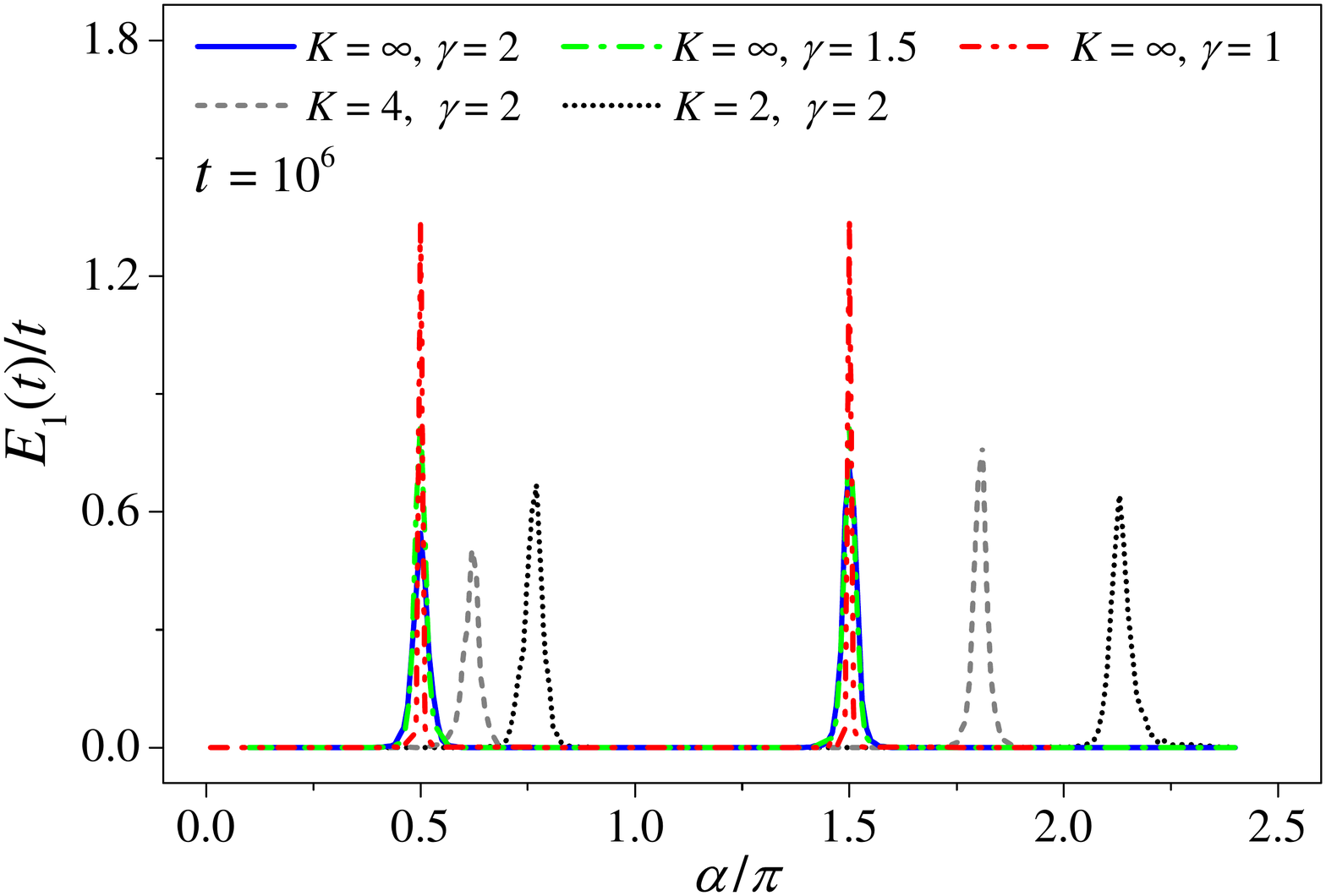}
\vskip-.5cm
\caption{The dynamical transitions exhibited by SM Model I evolve smoothly from the $\hbar_e$-driven IQHE exhibited by the spinful QKR. $\OM=(1,2\pi/\sqrt{5})$.}
\label{fig:3}
\end{figure}

Then we study how the $\alpha$-profile of the diffusion coefficient $\lim_{t\rightarrow\infty} E_1(t)/t$ depend on $K,\gamma$. For this purpose we choose a continuous contour in the parameter ($K,\gamma$) space along which the spinful QKR ($K=2,\gamma=2$) adopted in previous numerical experiments \cite{Tian14,Tian16} is deformed into the SM Model I ($K=+\infty,\gamma=1$). The contour consists of two pieces: (i) $\gamma$ is fixed to be $2$ while $K$ increases from $2$ to $+\infty$; then (ii) $K$ is fixed to be $+\infty$ while $\gamma$ decreases from $2$ to $1$. We compute the $\alpha$ profile of $E_1(t)/t$ at $t=10^6$
for different ($K,\gamma$) along the contour. Results are shown in Fig.~\ref{fig:3}. Obviously, along the piece (i) of the contour
every peak signalling the topological dynamical transition in the $\hbar_e$-driven IQHE is pushed continuously towards the origin and eventually the $m$th peak (counting from the left) arrives at $\alpha/\pi=m-\frac{1}{2}$ when $K=+\infty$; along the piece (ii) the peaks are pinned and their heights evolves continuously into those of the SM Model I. As such, though the SM models and the spinful QKR are very different,  and though  the supersymmetric field theory which was developed for the spinful QKR \cite{Tian14,Tian16} does not apply here, because of the absence of fast and slow mode separation \cite{Altland15}, the sequential dynamical localization--delocalization transitions exhibited by SM Model I evolve -- via the continuous change in ($K,\gamma$) along the contour -- smoothly from the $\hbar_e$-driven IQHE exhibited by the spinful QKR. Because in the latter a critical point separates two topologically distinct phases, whose topological angles differ by $2\pi$, the transition observed in SM Model I is of topological nature. However, it should be noted that the topological theory for the spinful QKR cannot be applied here, because of the lack of chaoticity in SM models.

\subsubsection{Stability of trajectories  of the equivalent classical dynamics}
\label{sec:instab}

\begin{figure}[t]
\vskip-.2cm\hskip-.5cm
\includegraphics[width=\columnwidth]{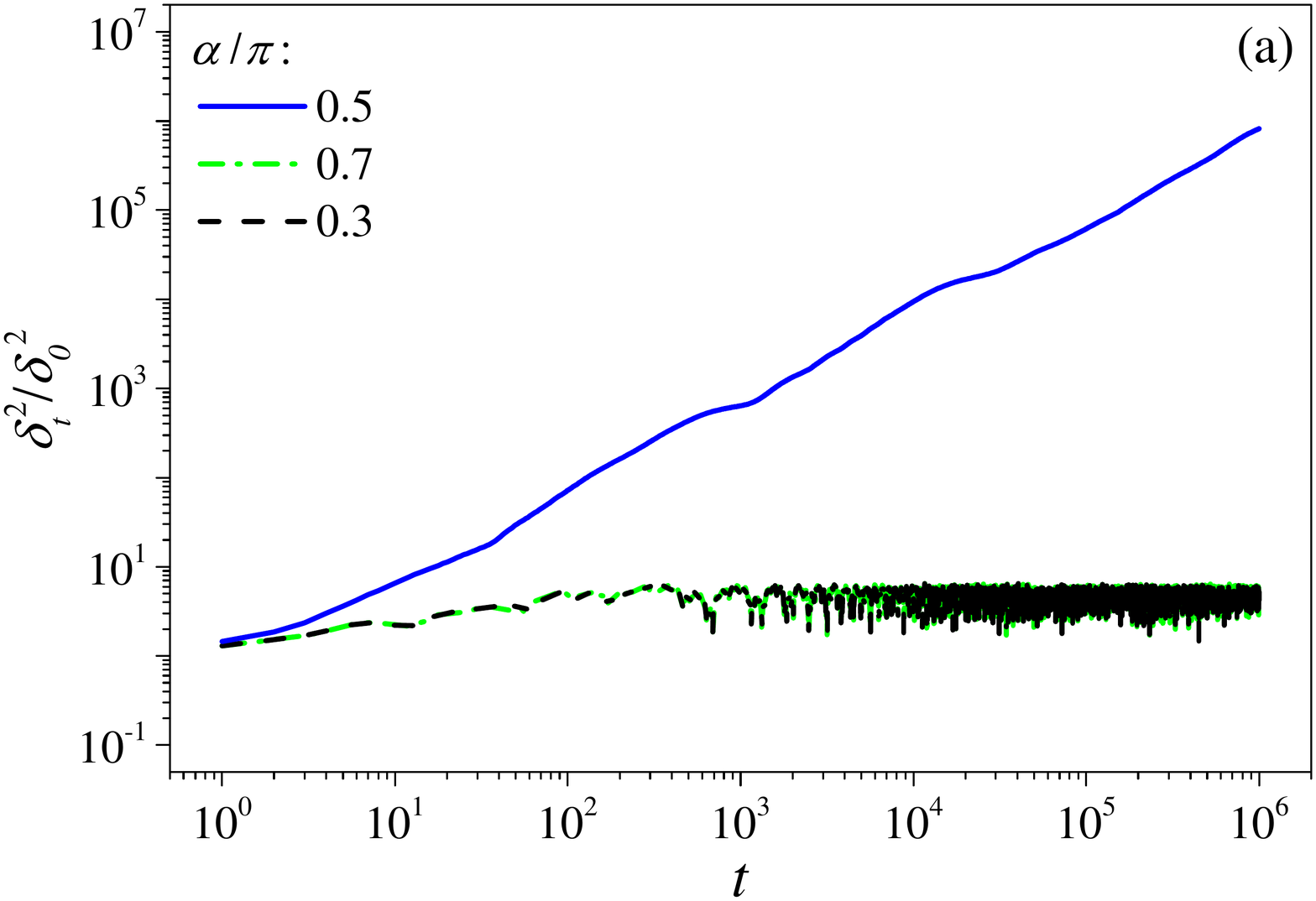}
\vskip-.3cm\hskip-.5cm
\includegraphics[width=\columnwidth]{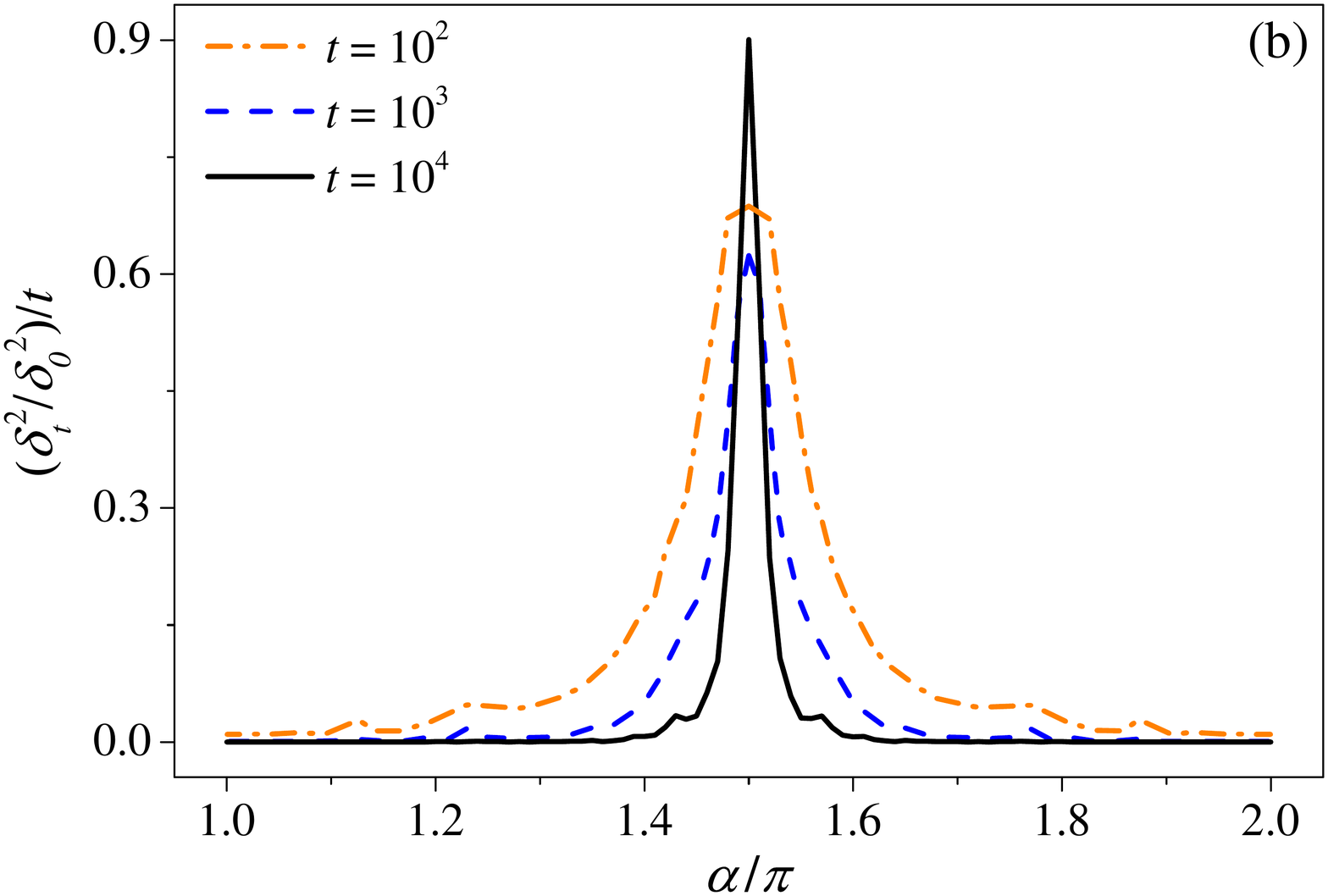}
\vskip-.5cm
\caption{Numerical experiments on SM Model I shows that the dynamical localization--delocalization transition shown in Fig.~\ref{fig:2} is translated into a change  in stability of
classical trajectories at half-integer $\alpha/\pi$. (a) The deviation of initially close trajectories grows unboundedly at half-integer $\alpha/\pi$
and it  saturates at long times away from these points. (b) As $t$ tends to infinity $(\delta_t/\delta_0)^2/t$ tends to a finite value at half-integer $\alpha/\pi$ while it tends to zero away from the critical point; $\OM=(1,2\pi/\sqrt{5})$.}
\label{fig:6}
\end{figure}

The equivalent classical dynamics described by Eq.~(\ref{eq:sysdin}) was numerically simulated up to $t=10^6$ for different values of $\alpha$. To probe the stability
of phase space trajectories we compute the time profile of the deviation $\delta_t$ of two trajectories which are initially close. Results are shown in Fig.~\ref{fig:6}. Time profiles of $(\delta_t/\delta_0)^2$ corresponding to different $\alpha$ are shown in (a), for which the average over initial conditions is performed, and the $\alpha$ profiles of $(\delta_t/\delta_0)^2/t$ at different long times are shown in (b). We see that, like $E_{1,2}(t)$, the long-time behavior of $(\delta_t/\delta_0)^2$ is very sensitive to the value of $\alpha/\pi$: for half-integer $\alpha/\pi$ it grows unboundedly, implying that the classical trajectory is unstable, whereas away from half-integer $\alpha/\pi$ it saturates at long times, implying that the classical trajectory is stable.

Moreover, for half-integer $\alpha/\pi$ the growth of $(\delta_t/\delta_0)^2$ is linear (a). Corresponding to this, as time increases the $\alpha$ profile of $(\delta_t/\delta_0)^2/t$ displays a sharp peak at half-integer $\alpha/\pi$ (b): $\lim_{t\to \infty}\frac{(\delta_t/\delta_0)^2}{t}$ is finite for half-integer $\alpha/\pi$ and vanishes otherwise. Therefore, the dynamical localization--delocalization transition  in terms of
quantum transport is translated into a transition in the stability of classical trajectories.
Remarkably, at the critical point the deviation  asymptotically grows approximately like $\sqrt{t}$, which is dramatically different from both the exponential instability of trajectories in dynamical chaos and the linear instability that is generic for Hamiltonian integrable systems \cite{linstab}.

\subsection{Dynamics of SM Model II}
\label{sec:nummodel2}

The same numerical procedures were repeated  with the constant in Eq.~(\ref{eq:sm2}) set to unity.
The observed dynamical behaviors of SM Model II are totally different from SM Model I. In particular, regardless of the value of $\alpha$, $E_1(t)$ always saturates at long times and the long-time diffusion coefficient $\lim_{t\rightarrow\infty}\frac{E_1(t)}{t}$ always vanishes (not shown). Thus, dynamical localization is always observed.


\subsection{Dynamics of SM Model III}
\label{sec:nummodel3}

The same numerical procedures were repeated for SM Model III with $d_3=p(\OT)(\cos(\theta_1)+\cos(\theta_2))$ and $d_3=0$, respectively. The
observed dynamics are totally opposite to those of SM Model II. In particular, dynamical delocalization is observed regardless of the value of $\alpha$,  with the one exception of the points $\alpha/\pi\in \mathbb{Z}$, for which $\U_{\alpha,\OM}=\pm\boldsymbol{T}_{\OM}$  leading to trivial localization.

\section{Spectral theory of SM models, and its dynamical implications}
\label{sec:math_results}

In this section, spectral theory is used to explore the
mechanisms of various quantum dynamical phenomena which were observed in numerical experiments on  SM models. Some rigorous mathematical results are presented, and their dynamical implications are discussed. To prevent mathematical technicalities from obscuring the overall physical picture, some proofs are shuffled to Sec.~\ref{sec:proof}.

\subsection{Spectral properties of general SM models}
\label{sec:gene_SM}

We first discuss the spectral properties of general SM models, without imposing any special features on the  $V_k(\OT)$. Such properties crucially depend  on arithmetic properties of the frequency vector $\OM$.
We always require $(\omega_1,\omega_2,\pi)$ to be an incommensurate triple. A general result about the spectrum of the Floquet operator
$\U_{\alpha,\OM}$ is stated below, and is proven  in Sec.\ref{sec:proof}.
Exact operator-theoretic terminology will be confined to those proofs, so here we fix a few terms of a more informal language.  The spectrum  of a unitary operator $\U$  in a Hilbert space $\cal H$ consists of complex numbers of the form $e^{i\chi}$ with $\chi\in\RM$, and it cannot be empty. The point spectrum of $\U$ is
that part of the spectrum which consists of proper eigenvalues, {\it i.e.},  of eigenvalues which are associated with normalizable eigenfunctions -- if there are any. Such eigenvalues are a countable set at most, and the corresponding  eigenfunctions span a subspace ${\cal H}_p$  of
$\cal H$,  which is invariant under $\U$; so its  orthogonal complement subspace ${\cal H}_c$ is also invariant under $\U$. If  ${\cal H}_p$ is not the whole of $\cal H$, then $\U$, restricted to ${\cal H}_c$, is still unitary, so it has a spectrum, which by construction cannot contain any proper eigenvalue. This is, in fact,  the continuous spectrum of $\U$,  and it can be informally pictured as the set  of generalized eigenvalues, which are associated with eigenfunctions that are not normalizable, because  they are, in some sense, extended.
The spectral type of $\U
$ is said to be pure, if either ${\cal H}_p$ or ${\cal H}_c$ is the whole $\cal H$; in the former case it is called pure point, and in the latter case it is called purely continuous. The set of all proper eigenvalues of $\U$ will be denoted $\eig(\U)$.

\begin{proposi}
\label{propo:specpure}
The spectral type of $\U_{\alpha,\OM}$ is pure. If it is pure point, then there is $\lambda\in[0,\pi]$ such that
   \begin{equation}
   \label{eq:specq}
   \eig\,(\U_{\alpha,\OM})\,=\,\left\{e^{i(\pm\lambda+\pmb{n}\pmb \cdot\pmb{\omega})}\right\}_{\pmb{n}\in\ZM^2}\;.
\end{equation}
In all cases, the spectrum is dense in the unit circle.
 \end{proposi}

\noindent{\bf Remarks}: (i) The point spectrum (\ref{eq:specq}) can be  degenerate only if  it contains the eigenvalue $e^{i\pmb p\pmb \cdot\pmb\omega/2+in\pi}$ for some $n\in\ZM$, $\pmb p\in\ZM^2$, and then it is doubly degenerate. (ii) The phase $\lambda$ is not uniquely defined, because the same set (\ref{eq:specq}) can be generated with infinitely many different choices of $\lambda$.

When a SM model is viewed in the classical picture, as described by Eq.~(\ref{eq:FP}), its spectral analysis is based on the Perron-Frobenius operator $\BU_{\alpha,\OM}$. The functions on the phase space $\Omega$ which depend only on $\OT$ are a closed subspace ${\mathfrak H}_0$
of $\mathfrak H$, that can be identified with $L^2(\TM^2,dm)$. This subspace is invariant under $\BU_{\alpha,\OM}$. In Sec.~\ref{sec:proof} we prove the following result:
\begin{proposi}
\label{propo:specc}
(i) The spectrum of $\BU_{\alpha,\OM}$ in ${\mathfrak H}_0$ is simple and pure point, with eigenphases ${\pmb n}\pmb{\cdot}{\pmb\omega}$ where ${\pmb n}\equiv (n_1,n_2)\in\ZM^2$. (ii)  If the spectrum of $\U_{\alpha,\OM}$ is pure point then the spectrum of $\BU_{\alpha,\OM}$ in  ${\mathfrak H}$ is pure point, and
   \begin{equation}
   \eig\,(\BU_{\alpha,\OM})\,=\,\left\{e^{i(m\lambda+{\pmb n}\pmb{\cdot}\pmb{\omega})}\right\}_{m,n_1,n_2\in\ZM}\;,
   \end{equation}
where the parameter $\lambda$ is the same as that in the spectrum (\ref{eq:specq}). (iii) If the spectrum of $\U_{\alpha,\OM}$ is continuous (hence purely continuous by Proposition \ref{propo:specpure}) then the spectrum of $\BU_{\alpha,\OM}$ in ${\mathfrak  H}$ has a continuous component.
\end{proposi}

Pure continuity of the spectrum of $\U_{\alpha,\OM}$ does not by necessity  entail pure continuity of the spectrum of $\BU_{\alpha,\OM}$ in the orthogonal complement ${\mathfrak H}_0^{\bot}$ of ${\mathfrak H}_0$;  the latter may still have a pure point component. This is somehow related to ergodicity of the classical description of SM models. Ergodicity is a capital notion in the theory of classical dynamical systems, and can be stated in purely spectral terms: $\CS_{\alpha,\OM}$ is ergodic, if and only if $1$ is a simple proper eigenvalue of $\BU_{\alpha,\OM}$ \cite{Sinai}.
 Since SM models bear a quantum--classical juxtaposition,
 a natural question arises, how would ergodicity in the classical description of SM models
 be mirrored  in the quantum description of SM models.
 The following result presents a partial answer. Loosely speaking, an operator in $\cal H$ is fibered, if it does not couple different position eigenstates; more properly, if it acts as in $\pmb\psi(\pmb\theta)\mapsto
 \boldsymbol{O}(\pmb\theta)\pmb\psi(\pmb\theta)$,
where $\boldsymbol{O}
$ is a map from $\TM^2$ to the $2\times 2$ matrices.

\begin{proposi}
\label{prop:ergo}
If the dynamical system $(\Omega, \CS_{\alpha,\OM})$ is ergodic, then the spectrum of $\U_{\alpha,\OM}$ is continuous and "fiberwise simple"; i.e., $\U_{\alpha,\OM}$ does not commute with any nontrivial fibered linear operator.
\end{proposi}
\noindent{\bf Remark}: This result cannot be reversed.

Though not necessary for the elaborations that follow, it can be shown that the spectral properties of the Floquet operator $\U_{\alpha,\OM}$ are related to the cohomology of the classical dynamical system $(\Omega, \CS_{\alpha,\OM})$; see Appendix \ref{sec:cohomology}.

\subsection{Unitary symmetry $\boldsymbol{A}$ and duality}
\label{sec:spec_SM}


We now focus on a special class of SM models, which are
exemplified by those
considered in numerical experiments.  A SM model belongs to
this class, if it is defined by maps $\boldsymbol{d}$ that enjoy the following special symmetries:
\begin{enumerate}[(i)]
\label{classa}
  \item $d_1(\OT)$ is odd wrt $\theta_1$ and even wrt $\theta_2$;
  \item $d_2(\OT)$ is even wrt $\theta_1$ and odd wrt $\theta_2$;
  \item $d_3(\OT)$ is even wrt both $\theta_1$ and $\theta_2$.
\end{enumerate}
It has been previously found
\cite{Tian14,Tian16} that such symmetries are crucial to the IQHE-like phenomenon in the spinful QKR. SM models endowed with such symmetries  will be termed class-$*$ models. The SM Models I, II and III, as given in Eqs.~(\ref{eq:sm1}), (\ref{eq:sm2}) and (\ref{eq:sm3}), are class-$*$ models.

A key role will be played by the following Proposition~\ref{propo:specsim}, that crucially rests on the above conditions (i), (ii) and (iii), and its corollaries.

\begin{proposi}
\label{propo:specsim}
For any class-$*$ model  a unitary symmetry $\pmb A$  \cite{note_unitary} exists, such that 
 \begin{equation}\label{eq:asymm}
 \pmb U^{\dagger}_{{\pi-\alpha,\pmb\omega}}\;=\;-\;\pmb A\;\pmb U_{{\alpha,\pmb\omega}}\;\pmb A\;,\,
 \pmb U^{\dagger}_{{2\pi-\alpha,\pmb\omega}}\;=\;\pmb A\;\pmb U_{{\alpha,\pmb\omega}}\;\pmb A\;.
 \end{equation}
\end{proposi}
{\it Proof}\,: The first identity in Eq.~(\ref{eq:asymm}) will be proven first; the proof of the second is essentially identical. First of all, from the general definition of $\boldsymbol{M}_\alpha$ it follows that
\begin{eqnarray}
\label{eq:symm}
  {\pmb M}_{2\pi-\alpha}={\pmb M}_{\alpha}^{\dagger},\,\,
  \pmb M_{\pi+\alpha}=-{\pmb M}_{\alpha},\,\,
  {\pmb M}_{\alpha}^{\dagger}=-
\pmb M_{\pi-\alpha},
\end{eqnarray}
which hold true independently of the symmetries of  $\boldsymbol{d}$. Let $\pmb R_{{\pmb\eta}}$ denote the reflection operator in a point $\pmb\eta\in\TM^2$: that is, $(\pmb R_{{\pmb\eta}}\pmb\psi)(\OT)=\pmb\psi(2\pmb\eta-\OT)$. From $\pmb U^{\dagger}_{\alpha,\OM}=\pmb T_{-\pmb\omega}\pmb M_{\alpha}^{\dagger}$ and the third identity in Eq.~(\ref{eq:symm}), we obtain that
\begin{equation}
\label{eq:pf1}
\pmb U^{\dagger}_{\alpha,\pmb\omega}\,=\,-\;\pmb T_{-\pmb\omega}\,\pmb U_{\pi-\alpha, -\pmb\omega}\,\pmb T_{\pmb\omega}.
\end{equation}
Thanks to the assumed symmetries (i), (ii) and (iii), the unitary operator $\pmb B=\pmb\sigma_3\pmb R_{\pmb 0}$
commutes with  $\pmb M_{\alpha}$, and $\pmb B\pmb T_{\pmb\omega}\pmb B^{\dagger}=\pmb T_{-\pmb\omega}$.
As a result,
\begin{equation}\label{eq:U_Bsymm}
  \pmb U_{\alpha, -\pmb\omega}\,=\,\pmb B\,\pmb U_{\alpha,\pmb\omega}\,\pmb B^{\dagger}\,.
\end{equation}
Applying this to the right-hand side
of Eq.~(\ref{eq:pf1}) gives
\begin{equation}\label{eq:U_Bsymm1}
\pmb U^{\dagger}_{\alpha,\pmb\omega}\,=\,-(\pmb T_{-\pmb\omega}\pmb B)\,\pmb U_{\pi-\alpha,\pmb\omega}\,(\pmb T_{-\pmb\omega}\pmb B)^{\dagger}\;.
\end{equation}
Therefore, the first identity in Eq.~(\ref{eq:asymm}) is true with $\boldsymbol{A}$ given by:
\begin{equation}
\label{eq:AB}
\pmb A\;=\;\pmb T_{-\pmb\omega}\pmb B\;=\;\pmb\sigma_3\pmb T_{-\pmb\omega}\pmb R_{\pmb 0}\;=\;\pmb\sigma_3\pmb R_{{-\pmb\omega/2}}\;\,\,\,.
\end{equation}
To prove the second identity in Eq.~(\ref{eq:asymm}), we use the first identity in Eq.~(\ref{eq:symm}) instead of the third. $\Box$\\

\noindent {\bf Remark}: Together with $\boldsymbol{A}^2=1$, Eq.~(\ref{eq:asymm}) implies, in particular, that the evolution operator $i\pmb U_{\alpha,\,\pmb\omega}$ at $\alpha=\pi/2$, $3\pi/2$, $0$ and $\pi$ are time-reversal invariant.
It is worth noting that also the associated classical map which is defined in Eq.~(\ref{eq:sysdin}) is time-reversal invariant;
the time-reversing involution is $(\OT,\,\pmb{\phi})\rightarrow(-\OT-\pmb\omega,\, -\pmb\sigma_3\pmb{\phi})$.

The symmetry $\boldsymbol{A}$  establishes a "duality" ~\cite{nature} between models whose $\alpha$ values in $[0,\pi]$ are symmetric with respect to $\pi/2$. In other words, the model at a value $\alpha_0$ of the parameter  $\alpha$ is dual to the model at $\alpha=\pi-\alpha_0$. The most immediate consequence of this duality is:
\begin{equation}\label{eq:spetral_relation}
  \eig(\pmb U_{{\pi-\alpha,\,\pmb\omega}})\;=\;-\;\eig(\pmb U_{{\alpha,\,\pmb\omega}})\,,
\end{equation}
because $\eig(\pmb U_{{\alpha,\,\pmb\omega}})$ is closed under complex conjugation, thanks to
the symmetry relation (\ref{eq:rholem}).

Models with
$\alpha=\pi/2$ are self-dual. This self-dual symmetry, {\it i.e.} the aforementioned time-reversal invariance
 has  nontrivial consequences:
\begin{coro}
\label{coro:topr}
If the operator $\pmb U_{\frac{\pi}{2},\,\pmb\omega}$ of a class-* model has a non-empty point spectrum, then there is $\pmb r\in\ZM^2$ so that
 \begin{equation}
 \label{eq:dich2}
 {\rm Eig}(\pmb U_{\frac{\pi}{2},\,\pmb\omega})\;=\;\bigl\{\pm i\;e^{i(\pmb n+\pmb r/2)\pmb \cdot\pmb\omega}\bigr\}_{\pmb n\in\ZM^2}\;.
 \end{equation}
  The corresponding eigenfunctions $\pmb{v}^{\pm}_{\pmb n}$ satisfy:
\begin{equation}
\label{eq:topeig}
\pmb A\pmb{v}^{\pm}_{\pmb n}(\OT)=c^{\pm}_{\pmb n}e^{i(2\pmb n+\pmb r)\pmb \cdot\OT}\pmb{v}^{\pm}_{\pmb n}(\OT)\;\;
\mbox{\rm with}\;\;(c^{\pm}_{\pmb n})^2=e^{i(2\pmb n+\pmb r)\pmb \cdot\pmb\omega}\;.
\end{equation}
\end{coro}
This corollary imposes quite demanding conditions on self-dual eigenfunctions (if any). In the case of SM Model III, they force the
 absence of proper eigenfunctions, so pure spectral continuity follows:
\begin{coro}
\label{coro:modIII}
The operators $\U_{\frac{\pi}{2},\,\pmb\omega}$ of SM Model III have purely continuous spectra.
\end{coro}
For SM Model I we cannot assess the spectral type of $\U_{\frac{\pi}{2},\,\pmb\omega}$. However the following  weaker result has crucial  dynamical consequences:
\begin{coro}
\label{coro:modI}
If $\U_{\frac{\pi}{2},\,\pmb\omega}$ of SM Model I has any proper eigenfunctions $\pmb u$, then such eigenfunctions are ``delocalized", in the sense that $E_{1,2}(\pmb u)=+\infty$, where $E_j$ are defined in Eq.~(\ref{eq:moment}).
\end{coro}
A proper mathematical statement should be: $\pmb u$ does not belong to
the domain of the momentum operators.

\subsection{Spectral properties vs dynamical delocalization}
\label{sec:spec_dyn}

The exact spectral results presented  above provide a basis for the study of  the dynamical delocalization which is observed in numerical experiments. It is important to point out that while such
exact results hold   for any incommensurate triple $(\omega_1,\omega_2,\pi)$,
the same is not true of other more detailed results to be discussed
in this subsection, which strongly rely on the value of
$\pmb\omega$ chosen as in Sec.~\ref{sec:trans}, and can reasonably be taken as representatives of a wide class of strongly incommensurate situations.

Based on the definition (\ref{eq:moment}) of $E_j$ ($j=1,2$) and $E$, we give  a precise definition of dynamical delocalization/localization in momentum space.
For given $\alpha,\OM$, the quantum dynamics $\ppsi_t=\U^t_{\alpha,\OM}\,\ppsi$ ($t\in \mathbb{Z}$) will be  said to exhibit
{\it dynamical delocalization} if for any $\ppsi$ such that $
E(\ppsi)<+\infty$  the following relation holds:
\begin{equation}
\label{eq:def_del}
  \limsup_{t\to\infty}\,
  E(\ppsi_t)\,=\,+\infty
\end{equation}
On the opposite, it will be said to exhibit {\it dynamical localization}  whenever  the right-hand side is finite.
It is important to note that  if $ E_j(\ppsi)<+\infty$ , then  $E_j(\ppsi_t)<+\infty$ at all times whenever the map $\pmb M_\alpha(\pmb\theta)$ that defines a SM model is smooth. This is not the case,  {\it e.g.} with SM Model III, where dynamical delocalization is somewhat trivial, as it will be shown below in Sec.~
\ref{sec:spec3}.
Next we recall that dynamical localization implies pure point spectra, and, equivalently, continuous spectra imply dynamical delocalization \cite{scho}.  So, for SM Model I (away from critical points) and Model II,
we have strong numerical evidence of pure point spectra, because  dynamical localization is clearly observed numerically. However, we have no mathematical proof (except, of course, in the trivial cases when
$\alpha=n\pi, n\in \mathbb{Z}
$).  Delocalization is produced by continuous spectra, but it  can also occur with pure point spectra, for  various reasons, {\it e.g.} non-uniform localization of proper eigenfunctions in momentum space, or else "delocalization" of eigenfunctions: meaning that proper
eigenfunctions, although  normalizable ({\it i.e.} not extended) do not yield a finite expectation value for
 the squared momentum operator. This follows from a general fact, which in our case can be stated as follows:
 \begin{proposi}
 \label{propo:meanerg}
 Let the spectrum of a unitary operator $\pmb U$ in $\cal H$ be simple and pure point, with delocalized  proper eigenfunctions. Then the quantum evolution  that is generated by $\pmb U$ in $\cal H$ is dynamically delocalized.
 \end{proposi}
A  compact proof of this standard result is given in Sec.~\ref{sec:proof}.
In view of Corollary \ref{coro:modI}, this may be the case with SM Model I at the critical point $\alpha=\pi/2$, which is self-dual.
So, even though we cannot exactly assess the spectral type of that model,  we can nevertheless state the following exact result, that provides a rigorous confirmation
for numerical observations:
\begin{proposi}
\label{final}
SM Model I is dynamically delocalized at the self-dual critical point $\alpha=\pi/2$.
\end{proposi}

{\it Proof}\,: Should the spectrum  of $\pmb U_{\frac{\pi}{2},\,\pmb\omega}$ be pure point, the claim  would immediately follow from Corollary \ref{coro:modI} and Proposition \ref{propo:meanerg}. The one possible alternative is pure
continuity, by spectral dichotomy. If that were the case,  then the claim would be trivial.
$\Box$\\

For SM Model III the situation is simpler:
\begin{proposi}
\label{proposi:spec3}
SM Model III is dynamically delocalized at all $\alpha\neq n\pi, n\in \mathbb{Z}$.   At least at the critical values of $\alpha$, the spectrum is purely continuous.
\end{proposi}
{\it Proof}\,: Delocalization follows from the fact that $\boldsymbol{M}_\alpha(\pmb\theta)$ exhibits points of essential discontinuity, {\it i.e.} points where it has infinitely many distinct limit values. Such are the points where the normalization factor $p(\pmb\theta)$ vanishes. Whatever the initial state   $\ppsi$, due to the very definition of $\U_{\alpha,\pmb\omega}$ such discontinuities are sooner or later transferred to $\ppsi_t$ whenever $\alpha\neq n\pi$, and they cannot be removed by re-defining $\ppsi_t$ in a set of measure zero. In contrast, a state for which $E_j$ has a finite value must be defined by functions which are, almost everywhere at least, equal to everywhere continuous functions.  Continuity of the spectrum is proven in Sec.~\ref{sec:spec3}.
$\Box$

This result receives intuitive support in the tight-binding-like formulation
of the SM model in Appendix \ref{sec:App}, because in the presence of discontinuities  the hopping terms in the lattice Hamiltonian (\ref{eq:hop})  quite slowly decay at long distances in momentum space.

\begin{figure}[t]
\vskip-.2cm\hskip-.5cm
\includegraphics[width=\columnwidth]{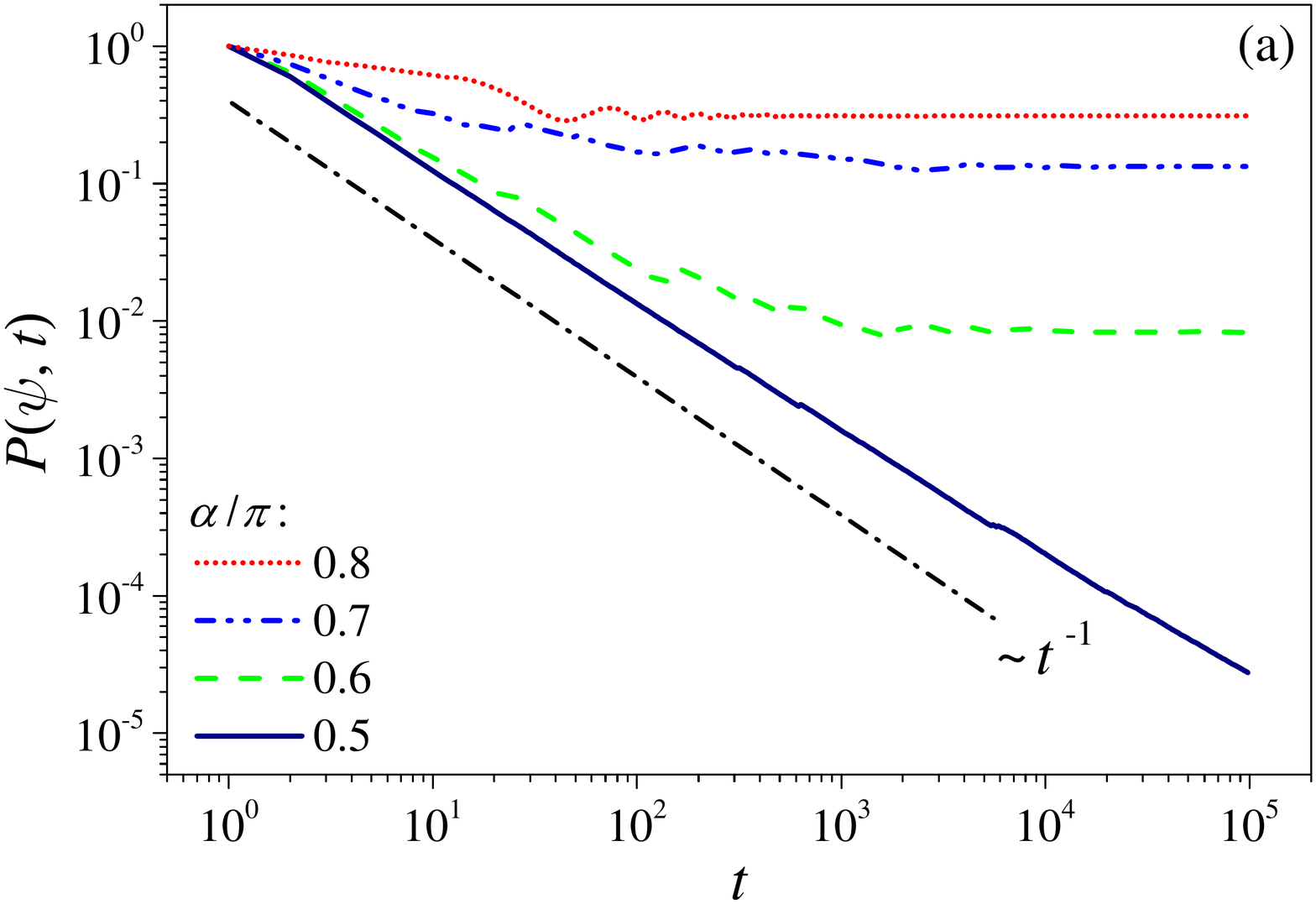}
\vskip-.2cm\hskip-.5cm
\includegraphics[width=\columnwidth]{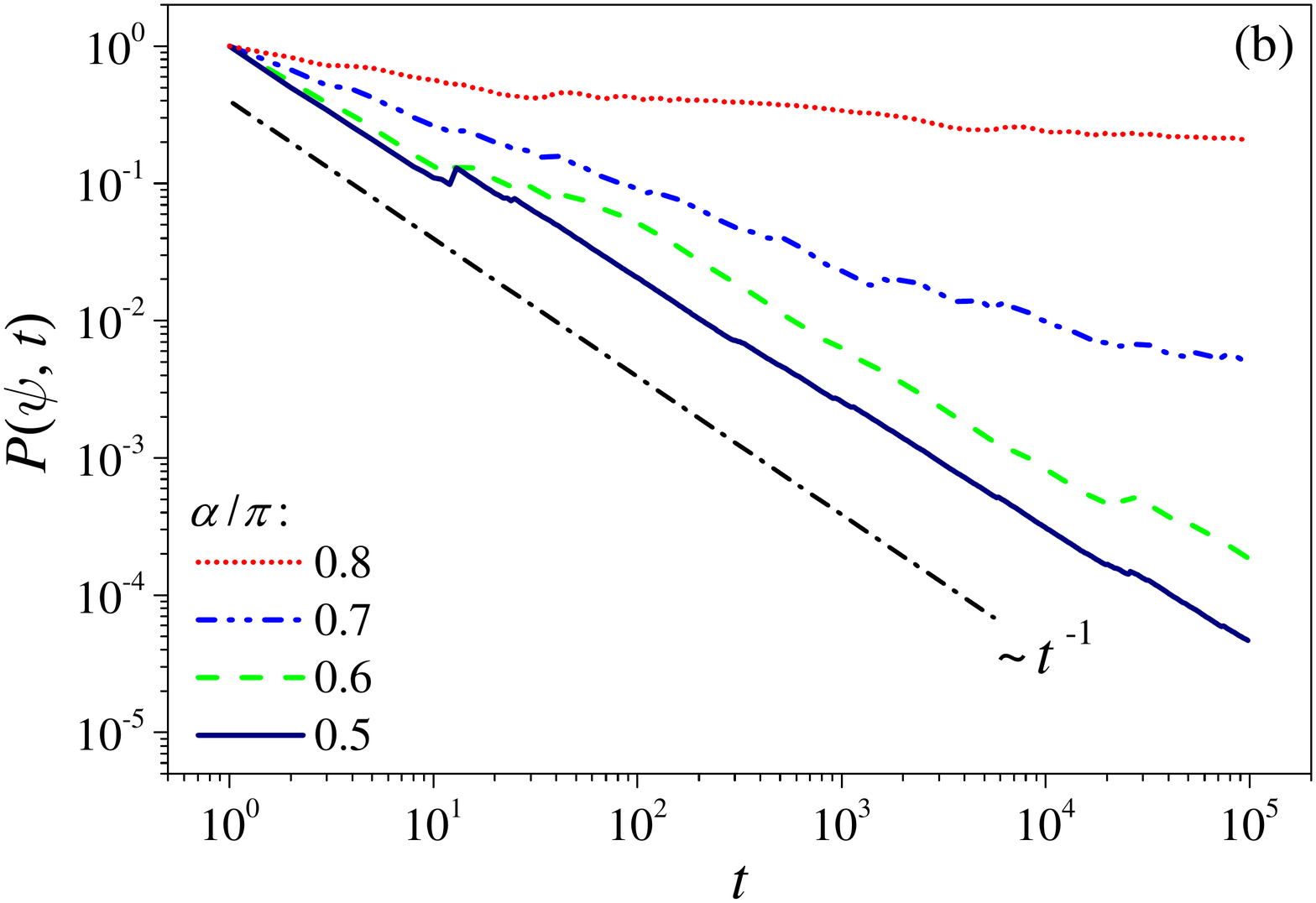}
\caption{$2$D simulation of the quantum evolution of an initial state $\ppsi$ localized at  $n_1=n_2=0$ for different $\alpha$. The results for the decay of the corresponding survival probabilities  for SM Models I and III (with $d_3=0$) are shown in (a) and (b), respectively. In both cases, $\OM=(2\pi x,2\pi x^2)$ (see text for the value of $x$).}
\label{fig:7}
\end{figure}

Though not proven,  continuity of the spectrum of SM Model I at the critical point is numerically supported, as follows.  For integer $t$ and $\pmb\psi\in\cal H$, let us define:
$$
P(\pmb\psi, t)\;\equiv\; \frac 1t\sum\limits_{s=0}^{t-1}|\langle\pmb\psi|\pmb\psi_s\rangle_{\cal H}|^2\;,
$$
where $\alpha,\pmb \omega$ are left understood, and  as usual $\pmb\psi_s=\U_{\alpha,\OM}^s\pmb\psi$. This  is the time-averaged probability of survival in the initial state $\pmb \psi$.  Wiener's theorem \cite{scho} states that
\begin{equation}
\label{wie}
\lim\limits_{t\to\infty}P(\pmb\psi, t)\;=\;
\sum\limits_{k}|\langle \pmb u_k|\pmb\psi\rangle_{\cal H}|^4\;,
\end{equation}
where the sum on the right-hand side is over all the proper eigenvectors $\pmb u_k$ of the Floquet operator $\U_{\alpha,\OM}$ (with an arbitrary numbering).
Vanishing of the limit in the left-hand side of Eq.~(\ref{wie}) is thus a necessary and sufficient condition for pure continuity of the spectrum.
 Numerical results of $2$D simulations for $P(\pmb\psi, t)$ are shown in Fig.~\ref{fig:7} for SM Models I and III, with $\pmb\psi$ chosen to be the $n_1=n_2= 0$ state, which is constant over $\mathbb{T}^2$.  As shown in (a), for SM Model I and noncritical $\alpha$, $P(\pmb\psi, t)$
 clearly stabilizes at large $t$, consistently with the pure point spectrum that is implied by  the observed dynamical localization. At critical $\alpha$, $P(\pmb\psi, t)$ appears to decay $\propto t^{-\nu}$ with $\nu$ equal or close to $1$ (a final slight deviation is probably
 due to finite-basis effects). Such  decay is almost identical to the one which  is observed for SM Model III at same $\alpha$
 [shown in (b)], where continuous spectrum is exactly proven. On such grounds spectral continuity of the critical model I is convincingly supported.

Next we address the subtler question, whether the spectrum of the critical model is absolutely continuous (a.c.) or singular continuous (s.c.). The latter type of spectrum is characterized by the fact that, despite continuity,  the spectral expansions of states $\pmb\psi\in\cal H$ concede positive (nonzero) weight to regions of the spectrum which have zero measure as subsets of the unit circle $\{e^{i\chi},\chi
\in\RM\}$. It is frequently associated with Cantor-like structures and multifractality. No exact analysis will be attempted here. However, the absence of  s.c. spectrum -- hence, pure absolute continuity of the  spectrum  --
 in the critical model I is numerically supported as follows.
 It is  an exact result \cite{dtwo} that the decay exponent $\nu
 $ of $P(\pmb\psi, t)$ is equal to the multifractal dimension
 $D_2$ of the local density of states (LDOS) associated with the state $\pmb\psi$; and $D_2=1$ is
 a strong indication \cite{footnote_ph}
 that the continuous spectrum is actually
 a.c..
    This appears to be the case with SM Model I at criticality,  because $D_2=\nu
 \approx 1$ according to Fig.~\ref{fig:7}.

\subsection{Quantum transport from a.c. spectrum}
\label{sec:spec_ac}

  We finally address quantum transport and show that numerical results are consistent with a.c. spectrum.
  Whenever $\pmb M_{\alpha}$ is a smooth function, the propagation over the $2$D $\pmb N$-lattice is subject to a {\it ballistic bound}; that is,   the asymptotic growth of $
  E(\pmb\psi_t)$
 cannot be faster than quadratic in time. This follows from Eq.~(\ref{eq:moment}),
 and is equivalent to  the fact  that classical trajectories can be linearly unstable at most [see the comments after Eq.~(\ref{eq:sqmod4})].
 The transport on the $1$D momentum lattice which is associated with the standard Maryland model and the QKR model are also subject to a ballistic bound. The following lower bound is valid for transport over discrete lattices of arbitrary dimension $D$
  induced by a unitary evolution group with a.c. spectrum \cite{scho,jmp}:
\begin{equation}
\label{lwbd}
  \frac 1t\sum\limits_{s=0}^{t-1}E(\pmb\psi_s)\;\geq\;\mbox{\rm const.}\,t^{\frac 2D}\;,\,\,\,(t\to\infty)\,.
    \end{equation}
 where $
 E(\pmb \psi)$ is the $D
 $-dimensional generalization of the definition in Eq.~(\ref{eq:moment}). With $D
 =1$, and  in the presence of a ballistic bound,  this inequality shows that a.c. spectra enforce ballistic transport \cite{footnote_bound}.
 Not so in higher dimension; for example,
 with $D
 =3$ it has been proved  \cite{JSP} that a.c. spectra can coexist with different diffusion exponents. In SM models the ballistic bound is attained, whenever
 $(\omega_1,\omega_2,\pi)$ is a commensurate triple. Then the quantum evolution enjoys translation invariance in momentum space, leading to band a.c. spectra and quadratic growth of $
 E(\pmb\psi_t)$. This scenario resembles
 the well-known "quantum resonances" of the QKR  \cite{qkres1,qkres2},  but
 the band structure is likely to be much more complex. This issue will not be
 investigated in the present paper, which is restricted to the incommensurate case.  In the strongly incommensurate situations which were numerically simulated, SM Model I at the critical value of $\alpha$  exhibits diffusive (or close to diffusive) transport, {\it i.e.}
 the long-time growth of $
 E(\pmb\psi_t)$ and $E_{1,2}(\pmb\psi_t)$ is linear in time. It is  not isotropic in momentum space, and the degree of anisotropy depends on the choice of $\pmb\omega$;  however, it should be noted that  diffusion  is not restricted to special directions.
 Diffusive growth is consistent with the bound (\ref{lwbd}) with $D
 =2$. The following heuristic argument indicates that indeed it is compatible with a.c. spectrum (in this case).
We start with the equation:
 \begin{equation}
 \label{specstab}
 E_j(\ppsi_{t})=
\bigl\|\,\sum_{s=0}^{t-1}\BU_{\alpha,\OM}^{s}H_j\,\bigr\|_{\mathfrak H}^2\,+\,\bigl\|\,\sum_{s=0}^{t-1}\BU_{\alpha,\OM}^{s}K_j\,\bigr\|^2_{\mathfrak H},
\end{equation}
which is derived in Sec.~\ref{sec:duality} exploiting the classical picture of SM models. The functions $H_j=H_j(\pmb\theta,\pmb\phi)$, $K_j=K_j(\pmb\theta,\pmb\phi)$ are defined on the phase space $\Omega$ of the classical model and are specified
immediately after Eq.~(\ref{eq:var_sum2}). Their explicit forms are
not important here, beyond the easily checked fact that they are
 orthogonal to the subspace of
the spin-independent functions, where $\BU_{\alpha,\OM}$ has a pure point spectrum according to Proposition
\ref{propo:specc}.
If the spectrum of $\pmb U_{\alpha,\OM}$ is a.c., then the spectrum of $\BU_{\alpha,\OM}$
has an a.c. component too (see the statement in the beginning of Sec.
~\ref{sec:prspecc}). Then a formal elaboration can be implemented. Let  $e^{i\chi}
$ denote generalized eigenvalues  of $\BU_{\alpha,\OM}$, and $|\chi,
\xi\rangle$ the corresponding generalized eigenvectors in Dirac notation; $\xi$ collectively denotes any additional quantum numbers. Then
\begin{equation}
\label{dirac}
\|\,\sum\limits_{s=0}^{t-1}\;\BU_{\alpha,\,\pmb\omega}^s H_j\,\|_{\mathfrak H}^2\;=\;
\int_{0}^{2\pi}d\chi
\,\frac{\sin^2\bigl(\frac{\chi
 t}{2}\bigr)}{\sin^2\bigl(\frac{\chi
}{2}\bigr)}\Lambda_{H_j}(\chi
)\,.
\end{equation}
where $\Lambda_{H_j}(\chi
)\equiv \sum_{\xi}|\langle\chi,
\xi|H_j\rangle|^2$ is the LDOS of $|H_j\rangle$ wrt
$\BU_{\alpha,\OM}$.
 Under the assumption of a.c. spectrum, this is an ordinary function
({\it i.e.}
 it is not Dirac $\delta$-like  or fractal-like), which will be further assumed to be smooth. A fully similar equation holds for the functions $K_j$.
The fraction under the integral sign in the right-hand side of Eq.~(\ref{dirac}), divided by $2\pi t$, in the limit $t\to\infty$ tends to a Dirac $\delta$. Then Eq.~(\ref{specstab}) shows  that the asymptotic growth of $
E(\pmb\psi_t)$ is linear in time. This result for SM models is not generic for quantum dynamics on $2$D lattices, as it rests on the classical
description of SM models which underlies Eq.~(\ref{specstab}).

\section{Equivalent classical description of quantum transport
}
\label{sec:duality}

In Sec.~\ref{sec:model_theoretical_tool} it was shown that
SM models can be viewed from either a ''quantum'' viewpoint  or from a "classical".  Combined with spectral theory, this offers advantages for exploring the SM dynamics from different perspectives. In particular, in this section
we show that the quantum transport in
momentum space
is translated into instability of trajectories $(\OT_t,\pmb{\phi}_t):=\CS_{\alpha,\OM}^t(\OT,\pmb{\phi})$ in the phase space $\Omega$. It is assumed throughout this section that $\pmb{M}_\alpha(\OT)$ is  everywhere  differentiable in $\mathbb{T}^2$.

\subsection{Instability of classical trajectories}
\label{sec:duality_gen}

The linear stability of a trajectory $(\OT_t,\pmb{\phi}_t)$ is determined by the behavior in time of the derivatives
$\partial_{\theta_j}\pmb{\phi}_t(\OT,\pmb{\phi})$ with $j=1,2$. From the classical dynamical equation (\ref{eq:sysdin}) it follows that
\begin{equation}
\label{eq:var}
\partial_{\theta_j}\pmb{\phi}_{t+1}\;=\;\pmb{M}_\alpha(\OT_{t+1})\partial_{\theta_j}\pmb{\phi}_t\;+\;(\partial_{\theta_j}\pmb{M}_\alpha(\OT_{t+1}))\;\pmb{\phi}_t\;.
\end{equation}
Define vectors $\pmb\xi_{j,t}\in\CM^2$ as follows:
\begin{equation}
\label{eq:xi}
\partial_{\theta_j}\pmb{\phi}_{t}\;=\;_{\rm T}\prod\limits_{s=0}^t\;\pmb{M}_\alpha(\OT_s)\;\pmb{\xi}_{j,t}\;,
\end{equation}
where $\pmb{\xi}_{j,0}=0$. Then by performing the summation: $\sum_{s=0}^{t-1}\partial_{\theta_j}\pmb{\phi}_{s}$, for which $\partial_{\theta_j}\pmb{\phi}_{s}$ is substituted by Eq.~(\ref{eq:xi}), we obtain that
\begin{equation}
\label{eq:var_sum1}
\pmb{\phi}_0\,\pmb{\cdot}\;\pmb{\xi}_{j,t}\;=\;\sum\limits_{s=0}^{t-1} H_j(\OT_s, \pmb{\phi}_s)\;
\end{equation}
and, similarly,
\begin{equation}
\label{eq:var_sum2}
\pmb{\rho}\pmb{\phi}_0\,\pmb{\cdot}\;\pmb{\xi}_{j,t}\;=\;\sum\limits_{s=0}^{t-1}K_j(\OT_s, \pmb{\phi}_s)\;,
\end{equation}
where $H_j(\OT,\pmb{\phi})=\pmb{\phi}\pmb{\cdot}\pmb{M}_\alpha^{\dagger}(\OT+\pmb{\omega})\partial_{\theta_j}\pmb{M}_\alpha(\OT+\pmb{\omega})\pmb{\phi}$ and $K_j(\OT,\pmb{\phi})=\pmb{\rho}\pmb{\phi}\pmb{\cdot}\pmb{M}_\alpha^{\dagger}(\OT+\pmb{\omega})\partial_{\theta_j}\pmb{M}_\alpha(\OT+\pmb\omega)\pmb\phi$. Because of $\pmb{\rho}\pmb{\phi}_0\pmb{\cdot}\pmb{\phi}_{0}=0$, taking squared moduli in both equations, and summing them together, we obtain that
\begin{equation}
\label{eq:preerg}
\pmb{\vert}\partial_{\theta_j}\pmb{\phi}_{t}\pmb{\vert}^2=\pmb{\vert}\pmb{\xi}_{j,t}\pmb{\vert}^2=
\bigl\vert\sum\limits_{s=0}^{t-1} H_j(\OT_s,\pmb\phi_s)\bigr\vert^2+\bigl\vert\sum\limits_{s=0}^{t-1} K_j(\OT_s,\pmb\phi_s)\bigr\vert^2,
\end{equation}
where $\pmb{\vert}\cdot\pmb{\vert}$ stands for the $\CM^2$ norm. As $H_j,K_j$ are by assumption bounded functions, this result shows that trajectories $(\OT_t,\pmb{\phi}_t)$ are, at worst, linearly unstable, in sharp contrast to the exponential instability of trajectories for dynamical chaos \cite{Chirikov79}.

To proceed we let
\begin{eqnarray}\label{eq:def_G}
  G_{\ppsi,t}(\OT,\pmb{\phi})&:=&(\BU_{\alpha,\OM}^{-t} G_{\ppsi})(\OT,\pmb{\phi})=G_{\ppsi}(\CS^t_{\alpha,\OM}(\OT,\pmb{\phi}))\nonumber\\
  &=&G_{\ppsi}(\OT_t,\pmb{\phi}_t)
\end{eqnarray}
with the function $G_{\ppsi}$ defined as in Eq.~(\ref{eq:def_F}). Taking the derivative wrt $\theta_j$ on both sides, we obtain that
\begin{equation}
\label{eq:Gj1}
\partial_{\theta_j}G_{\ppsi,t}(\OT,\pmb{\phi})\;=\;\ppsi(\OT_t)\pmb{\cdot}\partial_{\theta_j}\pmb{\phi}_t\;+\;
\partial_{\theta_j}\ppsi(\OT_t)\pmb{\cdot}\pmb{\phi}_t\;.
\end{equation}
On the other hand, thanks to Eq.~(\ref{eq:conn}) $G_{\ppsi,t}(\OT,\pmb{\phi})=\ppsi_{-t}(\OT)\pmb{\cdot}\pmb{\phi}$, where the initial spinor $\ppsi_{0}\equiv\ppsi$. Taking the derivative wrt $\theta_j$ on both sides of this equation entails
\begin{equation}
\label{eq:Gj2}
\partial_{\theta_j}G_{\ppsi,t}(\OT,\pmb{\phi})\;=\;\partial_{\theta_j}\ppsi_{-t}(\OT)\,\pmb{\cdot}\,\pmb{\phi}\;.
\end{equation}
Choosing the initial spinor $\ppsi$ to be independent of $\OT$ and equating Eqs.~(\ref{eq:Gj1}) and (\ref{eq:Gj2}), we obtain that
\begin{equation}
\label{eq:sqmod1}
\partial_{\theta_j}\pmb{\phi}_t(\OT,\pmb{\phi})\pmb{\cdot}\ppsi\;=\;\pmb{\phi}\pmb{\cdot}\partial_{\theta_j} \ppsi_{-t}(\OT)\;.
\end{equation}
Replacing $\ppsi$ by $\pmb{\rho}\,\ppsi$, a short calculation shows that
\begin{equation}
\label{eq:sqmod2}
\partial_{\theta_j}\pmb{\phi}_t(\OT,\pmb{\phi})\pmb{\cdot}\pmb{\rho}\ppsi\;=\;
-\partial_{\theta_j} \ppsi_{-t}(\OT)\pmb{\cdot}\pmb{\rho}\pmb{\phi}.
\end{equation}
Summing squared moduli in the last two equations yields
\begin{equation}
\label{eq:sqmod3}
\pmb{\vert}\partial_{\theta_j}\pmb{\phi}_t(\OT,\pmb{\phi})\pmb{\vert}^2\;=\;\pmb{\vert}\partial_{\theta_j} \ppsi_{-t}(\OT)\pmb{\vert}^2\;.
\end{equation}
This identity has the  important consequence, that neither the left-hand side nor the right-hand side depends on the choice of $\pmb{\phi}$ and of the initial constant spinor $\pmb{\psi}$.

\subsection{Derivation of Eq.~(\ref{specstab})}
\label{sec:duality_der}

Combining Eq.~(\ref{eq:sqmod3}) with Eq.~(\ref{eq:moment}), we obtain:
\begin{eqnarray}
\label{eq:Ejt}
  E_j(\ppsi_{-t})&=&\int_{\TM^2}dm(\OT)\;\pmb{\vert}\partial_{\theta_j}\pmb{\phi}_t(\OT,\pmb{\phi})\pmb{\vert}^2\nonumber\\
&=&\int_{\Omega}d\mu(\OT,\pmb{\phi})\;\pmb{\vert}\partial_{\theta_j}\pmb{\phi}_t(\OT,\pmb{\phi})\pmb{\vert}^2\,,
\end{eqnarray}
where the second line is obtained by
performing an additional integral wrt $\pmb{\phi}$, which is legitimate because, as noted above, the integrand is independent of $\pmb{\phi}$. With the substitution of Eq.~(\ref{eq:preerg}), Eq.~(\ref{eq:Ejt}) reduces to
\begin{eqnarray}
\label{eq:sqmod4}
E_j(\ppsi_{-t})=
\bigl\|\,\sum_{s=0}^{t-1}\BU_{\alpha,\OM}^{-s}H_j\,\bigr\|_{\mathfrak H}^2\,+\,\bigl\|\,\sum_{s=0}^{t-1}\BU_{\alpha,\OM}^{-s}K_j\,\bigr\|^2_{\mathfrak H}\,\,\,
\end{eqnarray}
or equivalently, Eq.~(\ref{specstab}).
Equation (\ref{eq:Ejt}) establishes the announced juxtaposition
between quantum transport and instability of classical trajectories. Moreover, from Eq.~(\ref{eq:sqmod4}) or (\ref{specstab})
it follows that, for the SM model, the quantum transport in momentum space is never faster than ballistic. It is worth mentioning  that if the equivalent classical dynamical systems of SM models were known to be ergodic then Eq.~(\ref{eq:sqmod4}) or (\ref{specstab})
would entail sub-ballistic transport, because the phase space averages  of both $H_j$ and $K_j$ vanish, and the same would be true for their  time averages.

\section{Proofs of some mathematical results}
\label{sec:proof}

Here we present proofs of some mathematical results presented in Sec.~\ref{sec:math_results}.
The readers who are not interested in mathematical details can skip this section.

\subsection{Proof of Proposition \ref{propo:specpure} }
\label{sec:prspecpure}

Here an extended version of  Proposition \ref{propo:specpure} will be proven. It includes the statement, that if the spectrum is continuous, then it is either purely absolutely continuous, or purely singular continuous.

(i) Let $\pmb{u}(\OT)\in{\cal H}_1:=\{\pmb{\psi}\in{\cal H}: \pmb{\psi}(\OT)\in\CM^2_1\;,\forall\OT\}$, and for any ${\pmb n}\in\ZM^2$  define vectors ${\pmb{u}}_{+,\pmb n}:=e^{i{\pmb n}\pmb \cdot{\OT}}{\pmb{u}}$ and ${\pmb{u}}_{-,\pmb n}:=\RH {\pmb{u}}_{+,\pmb n}=e^{-i{\pmb n}\pmb \cdot{\OT}}\RH\,{\pmb{u}}$. Then the vectors $\{{\pmb{u}}_{\pm,\pmb n}\}_{\pmb n\in \mathbb{Z}^2}$ are a {\it total set} of vectors in $\cal H$. To see this, let $\Psi\in{\cal H}$ be orthogonal to all vectors ${\pmb{u}}_{\pm,\pmb n}$. In particular,
\begin{equation}\label{eq:totset1}
  \int_{{\mathbb T}^2}dm(\OT)\;e^{i{\pmb n}\pmb \cdot{\OT}}\;\Psi(\OT)\pmb \cdot {\pmb{u}}(\OT)\;=\;0,\,\;\forall
 {\pmb n}\in\ZM^2\;.
\end{equation}
As $\pmb{u}(\OT)$ never vanishes, completeness of the Fourier basis entails that $\Psi(\OT)$ is almost everywhere in ${\mathbb T}^2$ orthogonal
 to $\pmb{u}(\OT)$. Hence $\Psi$ is a multiple of $\RH\, \pmb{u}$, because the latter vector, together with $\pmb{u}$, makes a basis for $\CM^2$. So $\Psi=c(\OT)\RH\, \pmb{u}$ where $c(\OT)$ is some measurable function. By the same argument, one has
\begin{equation}\label{eq:totset2}
  \int_{{\mathbb T}^2}dm(\OT)\;e^{-i{\pmb n}\pmb \cdot{\OT}}\;\Psi(\OT)\pmb \cdot\RH{\pmb{u}}(\OT)\;=\;0,\,\;\forall
 {\pmb n}\in\ZM^2\;,
\end{equation}
from which it follows that $\Psi$ must also be a scalar multiple of $\pmb{u}$: this is possible if and only if $c(\OT)\equiv 0$. Thus Eqs.~(\ref{eq:totset1}) and (\ref{eq:totset2}) can be satisfied simultaneously, if and only if $\Psi(\OT)\equiv 0$ .

Next, for any $t\in\ZM$, consider the correlations:
\begin{equation}\label{eq:corr}
  {\cal R}_{\pm,\pmb n}(t)\;=\;\langle{\pmb{u}}_{\pm,\pmb n}\;\vert\;\U_{\alpha,\OM}^t{\pmb{u}}_{\pm,\pmb n}\rangle_{\mathcal H}\;.
\end{equation}
It is a basic notion in spectral theory \cite{Simon1}  that such correlations are Fourier transforms of the spectral measures of the vectors ${\pmb{u}}_{\pm,\pmb n}$ (also known as
LDOS). From the definition of $\U_{\alpha,\OM}$ and from Eq.~(\ref{eq:rholem}), it follows that
\begin{equation}\label{eq:spec_R}
  {\cal R}_{+,\pmb n}(t)\;=\;{\cal R}_{-,\pmb n}^*(t)\;=\;e^{-i{\pmb n}\pmb \cdot{\pmb\omega}t}{\cal R}_{+,\pmb 0}(t).
\end{equation}
This shows that the spectral measures of all vectors ${\pmb{u}}_{\pm,\pmb n}$  can be  obtained from the spectral measure of any single one of them, say ${\pmb{u}}_{+,\pmb 0}$,  by irrational rotations and conjugation in the unit circle. As these two operations cannot change the nature of the spectral measure (pure point, absolutely or singular continuous), and $\{{\pmb{u}}_{\pm,\pmb n}\}_{\pmb n\in \mathbb{Z}^2}$ is a total set, purity of the spectrum follows. Furthermore, since the set  $\{e^{i{\pmb n}\pmb \cdot{\pmb\omega}}\}_{\pmb n\in \mathbb{Z}^2}$ is dense in $\mathbb{C}_1$,
the closure of the spectrum is always the full unit circle.

(ii) If the spectrum is pure point, then let $\pmb{u}(\OT)$ be an eigenvector, with eigenvalue $e^{i\lambda}$. Its $\CM^2$ norm is constant, as it can be seen on taking norms of both sides of $\boldsymbol{M}_\alpha \pmb{u}(\OT-\pmb\omega)=e^{i\lambda}\pmb{u}(\OT)$ and then using ergodicity of the $\pmb\omega$-shift $\boldsymbol{\tau_\OM}$. From the definition of $\U_{\alpha,\OM}$, it immediately follows that $\pmb{u}_{\pm,\pmb n}$ is an eigenvector, with eigenvalue $e^{\pm i\lambda\mp i\pmb n\pmb \cdot\pmb\omega}$. Since $\{{\pmb{u}}_{\pm,\pmb n}\}_{\pmb n\in \mathbb{Z}^2}$ is a total set, Eq.~(\ref{eq:specq}) follows.
$\Box$

\subsection{Proof of Proposition \ref{propo:specc}}
\label{sec:prspecc}
Here an extended version of the Proposition is proven. It includes a proof that the continuous component of the spectrum of $\BU_{\alpha,\OM}$ shares the type (s.c. or a.c.)
of the spectrum  of $\U_{\alpha,\OM}$. The functions $e^{i\pmb{n}\pmb{\cdot}\pmb{\theta}}$, (${\pmb n}\in\ZM^2$) are a complete basis of $\mathfrak{H}_0$, so claim (i) follows from
$\BU_{\alpha,\OM}e^{i\pmb{n}\pmb{\cdot}\pmb{\theta}}=e^{-i\pmb{n}\pmb{\cdot}\pmb{\omega}}e^{i\pmb{n}\pmb{\cdot}
\pmb{\theta}}$.

Next, let $p,q$ be nonnegative integers, $j,k=\pm 1$, and $\ppsi(\pmb{\theta})\in\cal H$. 
The functions
\begin{gather}
F_{\pmb{n},j,k,p,q,\ppsi}(\pmb{\theta},\pphi)\;=
\;e^{i{\pmb{n}}\pmb{\cdot}\pmb{\theta}}\;[\ppsi^{(j)}(\pmb{\theta})\pmb{\cdot}\pphi]^p[\pphi\pmb{\cdot}\ppsi^{(k)}(\pmb{\theta})]^q
\label{eq:cas}
\end{gather}
with $\ppsi^{(-1)}=\ppsi$, $\ppsi^{(+1)}=\pmb{\rho}\,\ppsi$,
$p+q\neq 0$, are a total set in ${\mathfrak H}_0^{\bot}$ whenever $\ppsi(\pmb{\theta})$ is almost everywhere nonzero; this directly follows, {\it e.g.} from the complex Stone-Weierstrass theorem \cite{Simon1}.
Next, thanks to  Eq.~(\ref{eq:conn}) we note that
\begin{eqnarray}
\label{eq:uqcq}
\BU_{\alpha,\OM}(F_{\pmb{n},j,k,p,q,\ppsi})\;=\;e^{-i\pmb{n}\pmb{\cdot}\pmb{\omega}}F_{\pmb{n},j,k
,p,q,\,\ppsi'}\,,
\end{eqnarray}
where $\ppsi'=\U_{\alpha,\OM}\ppsi$.

On one branch of the dichotomy, the spectrum of $\U_{\alpha,\OM}$ is pure point. Let $\ppsi$ be the eigenfunction of $\U_{\alpha,\OM}$ with eigenvalue $e^{i\lambda}$. From Eq.~(\ref{eq:uqcq}) it then follows that $F_{\pmb{n},j,k,p,q,\ppsi}$ is the eigenfunction of $\BU_{\alpha,\OM}$ with eigenvalue $e^{i[\lambda(jp-kq)-\pmb{n}\pmb{\cdot}\pmb{\omega}]}$. This holds true for all $\pmb{n},j,k, p,q$; moreover, eigenfunctions of $\U_{\alpha,\OM}$ are almost everywhere nonzero (owing to  ergodicity of $\OM$-shift in ${\mathbb T}^2$), so the pure point spectral subspace of $\BU_{\alpha,\OM}$ restricted on $\mathfrak{H}^\perp_0$ includes a total set. Thus the spectrum of $\BU_{\alpha,\OM}$ in $\mathfrak{H}^\perp_0$ is pure point also. In combination with (i) this leads to (ii).

On the other branch of the dichotomy, the spectrum of $\U_{\alpha,\OM}$ is purely continuous. Denote
\begin{equation}
\label{eq:gfunc}
G_{\ppsi}(\OT,\pphi)=
\ppsi(\OT)\pmb{\cdot}\pphi\,,
\end{equation}
which is the function (\ref{eq:cas}) with $\pmb{n}=0, j=-1,p=1,q=0$. Then for any $t\in \mathbb{Z}$:
\begin{eqnarray}
\label{eq:spec_G}
\langle G_{\ppsi}\,|\,\BU^t_{\alpha,\OM} G_{\ppsi}\,\rangle_{\mathfrak H} &=&\int_{\Omega}d\mu(\OT,\pphi)\;[\ppsi(\OT)\pmb{\cdot}\pphi][\pphi\pmb{\cdot} \U^t_{\alpha,\OM}\ppsi(\OT)]\nonumber\\
&=&\tfrac12\int_{\TM^2} dm(\OT)\;\ppsi(\OT)\pmb{\cdot}\U^t_{\alpha,\OM}\ppsi(\OT)\nonumber\\
&=&
\tfrac12\langle
\ppsi\,|\,\U^t_{\alpha,\OM}\ppsi\rangle_{\mathcal H}\,.
\end{eqnarray}
Therefore, the spectral measure of $G$ wrt $\BU_{\alpha,\OM}$ is -- apart from normalization -- the same as the spectral measure
of $\ppsi$ wrt $\U_{\alpha,\OM}$. So (iii) and its extended version, claimed at the beginning of this subsection, follow. $\Box$

\subsection{Proof of Proposition \ref{prop:ergo}}
\label{sec:pr_ergo}

It will be shown  that if $\U_{\alpha,\OM}$ has a proper
eigenfunction $\pmb{u}$, and (or) $\U_{\alpha,\OM}$ commutes with a nontrivial (linear) fibered operator
$\pmb{{\cal B}}$, then
$(\Omega, \CS_{\alpha,\OM})$ is not ergodic.

By the very definition of ergodicity, $(\Omega, \CS_{\alpha,\OM})$ cannot be ergodic, if invariant functions exist, which are not almost everywhere constant. If $\U_{\alpha,\OM}$ has an
eigenfunction $\pmb{u}$, then $F({\OT},\pmb{u})=|\pmb{u}(\OT)\pmb{\cdot}\pmb{\phi}|$ is a nonconstant invariant function on $\Omega$. On the other hand, if $\U_{\alpha,\OM}$ commutes with a fibered operator
$\pmb{{\cal B}}$ then the following ``covariance relation'' holds:
\begin{equation}\label{eq:cov_rel}
\pmb{{\cal B}}(\OT) =\pmb{M}_\alpha(\OT)\,
\pmb{{\cal B}}(\OT-\OM)\,
 \pmb{M}_\alpha(\OT)^{-1}\;,\,\,\,\forall\OT\in\TM^2\;.
\end{equation}
Then a simple calculation shows that the function that is defined on $\Omega$ by $F(\OT, \pmb{\phi})=
\pmb{{\cal B}}(\OT)\pmb{\phi}\pmb{\cdot}\pmb{\phi}$ is not constant (because $
\pmb{{\cal B}}$ is nontrivial), yet it is invariant under the map $\CS_{\alpha,\OM}$. $\Box$

\subsection{Proof of Corollary \ref{coro:topr}}
\label{sec:pr_coro_topr}

The proof consists of two parts.

(i) Let $\pmb{u}$ be an eigenvector of $\U_{{\alpha,\pmb\omega}}$, and $e^{i\chi}$ the corresponding eigenvalue. Thanks to the first identity in Eq.~(\ref{eq:asymm}),   $\boldsymbol{A} \pmb{u}$ is an eigenvector of $\U_
 {{\pi-\alpha,\pmb\omega}}$, with eigenvalue $e^{-i\chi+i\pi}$.
 Since $(\omega_1,\omega_2,\pi)$ is an incommensurate triple, it easily follows from Eq.~(\ref{eq:specq}) that, if
 ${\rm Eig}(\U_{\frac{\pi}{2},\pmb\omega})$ is not empty, then  there are some $\boldsymbol{p},\boldsymbol{q}\in\mathbb{Z}^2$ so that $\chi=\lambda+\boldsymbol{p}\pmb \cdot\OM$ and $-\chi+\pi=\lambda+\boldsymbol{q}\pmb \cdot\OM$. Adding these two equalities together gives $\pi=2\lambda+(\boldsymbol{p}+\boldsymbol{q})\pmb \cdot\OM$. Thus there is some $\pmb r \in\ZM^2$ so that $\lambda$ in Eq.~(\ref{eq:specq}) can be expressed as $\lambda=\pi/2+\pmb r\pmb \cdot\pmb\omega/2$. Thus the spectrum is as declared in Eq.~(\ref{eq:dich2}) and it is not degenerate, see remark (i) following Proposition \ref{propo:specpure}.

(ii) Let $\pmb{v}^{\pm}_{\pmb n}$ be an eigenvector of $\U_{{\frac{\pi}{2},\pmb\omega}}$ that corresponds to the eigenvalue
 $\pm ie^{i(\pmb n+\pmb r/2)\pmb \cdot\pmb\omega}$; then thanks to the first identity in Eq.~(\ref{eq:asymm}), $\boldsymbol{A}\pmb{v}^{\pm}_{\boldsymbol{n}}$ is an eigenvector that corresponds to the eigenvalue $\pm ie^{-i(\pmb n+\pmb r/2)\pmb \cdot\pmb\omega}$. On the other hand, by the following relation:
\begin{eqnarray}
\label{eq:Ueig}
  &&\U_{\frac{\pi}{2},\OM}\pmb{v}_{\pmb n}^\pm (\OT)e^{i(2\pmb n+\pmb r)\pmb \cdot \OT}
=\boldsymbol{M}_{\frac{\pi}{2}}\pmb{v}_{\pmb n}^\pm (\OT-\OM)e^{i(2\pmb n+\pmb r)\pmb \cdot (\OT-\OM)}\nonumber\\
  &=&\pm ie^{i(\pmb n+\pmb r/2)\pmb \cdot \OM}\pmb{v}_{\pmb n}^\pm (\OT)e^{i(2\pmb n+\pmb r)\pmb \cdot (\OT-\OM)}\nonumber\\
  &=&\pm ie^{-i(\pmb n+\pmb r/2)\pmb \cdot \OM}\pmb{v}_{\pmb n}^\pm (\OT)e^{i(2\pmb n+\pmb r)\pmb \cdot \OT}
\end{eqnarray}
one finds that $\pmb{v}^{\pm}_{\pmb n}e^{i(2\pmb n+\pmb r)\pmb \cdot\OT}$ is an eigenvector that has the same eigenvalue as $\boldsymbol{A}\pmb{v}^{\pm}_{\boldsymbol{n}}$. Since, as it was noted  above, the  spectrum is not degenerate, the two eigenvectors must coincide up to a constant phase factor $c$.  Equation (\ref{eq:topeig}) directly follows. The statement about the factor  $c=c^{\pm}_{\pmb n}$ in Eq.~(\ref{eq:topeig}) is obtained by operating with $\boldsymbol{A}$ on both sides of the first equation in (\ref{eq:topeig}),
and thereafter using the same Eq.~(\ref{eq:topeig}) in the thus obtained equation. $\Box$

\subsection{Proof of Corollary \ref{coro:modIII}
}
\label{sec:spec3}

The operator $\U_{\frac{\pi}{2},\OM}$ of SM Model III is denoted by $\mathbb U$. We shall prove that the spectrum of $\mathbb{U}$ is continuous,
first with
$d_3=0$, and then with $d_3=\cos\theta_1+\cos\theta_2$.

(i) $d_3=0$. The corresponding operator $\pmb M_{\frac{\pi}{2}}$ is
\begin{eqnarray}
\label{eq:M}
{\pmb M}_{\frac{\pi}{2}}=\left(
  \begin{array}{cc}
    0 & -ie^{-i\Phi(\boldsymbol{\theta})} \\
    -ie^{i\Phi(\boldsymbol{\theta})} & 0 \\
  \end{array}
\right)=:{\pmb{\mathbb M}},
\end{eqnarray}
where $\Phi(\boldsymbol{\theta})$ is the argument of the complex number $\sin(\theta_1)+i\sin(\theta_2)$. Continuity  of the spectrum of
$\mathbb U$ will be proven by contradiction. Assume the spectrum to be pure point. Due to  Eq.~(\ref{eq:dich2}), $\mathbb U$ has an eigenvalue $ie^{i\pmb r\pmb\cdot\pmb\omega/2
}$ with a corresponding eigenfunction denoted by $\pmb{u}=(u_1,u_2)^T=\pmb{v}^+_{\pmb 0}$ ($T$ denotes the transpose). The latter satisfies
\begin{eqnarray}
\forall\, \OT\in \mathbb{T}^2:
e^{i\pmb{r}\pmb{\cdot}\pmb{\omega}/2} \pmb{u}(\OT)=-i\;\pmb{\mathbb M}(\OT)\pmb{u}(\OT-\pmb{\omega})\label{eq:u1}\\
\Leftrightarrow
\left\{\begin{array}{c}
         e^{i\pmb{r}\pmb{\cdot}\pmb{\omega}/2} u_1(\OT)=-i\;\pmb{\mathbb{M}}_{12}(\OT)u_2(\OT-\pmb{\omega}) \\
         e^{i\pmb{r}\pmb{\cdot}\pmb{\omega}/2} u_2(\OT)=-i\;\pmb{\mathbb{M}}_{21}(\OT)u_1(\OT-\pmb{\omega})
       \end{array}
\right..\quad
\label{eq:u2}
\end{eqnarray}
So from $\pmb{\mathbb M}_{12}\pmb{\mathbb M}_{21}=-1$ it follows that
\begin{equation}\label{eq:u1u2}
  u_1(\OT)u_2(\OT)=e^{-i\pmb r\pmb\cdot\pmb\omega}u_1(\OT-\pmb\omega)u_2(\OT-\pmb\omega).
\end{equation}
This entails
\begin{equation}\label{eq:u1u2sol}
  u_1(\OT)u_2(\OT)=c_1e^{-i\pmb r\pmb\cdot\OT},
\end{equation}
where $c_1\neq 0$ is a constant.

On the other hand, from Eq.~(\ref{eq:topeig}) it follows that
\begin{equation}\label{eq:uu2}
  u_2(-\OM-\OT)=-e^{i\pmb r\pmb\cdot\pmb\omega/2}e^{i\pmb r\pmb\cdot\OT}u_2(\OT).
\end{equation}
Combining this with the first equation in (\ref{eq:u2}) gives
\begin{equation}\label{eq:uu1}
  u_2(\OT)=-e^{-i\pmb r\pmb\cdot\OT}\pmb{\mathbb M}_{21}(\OT) u_1(-\OT).
\end{equation}
With this substitution Eq.~(\ref{eq:u1u2sol}) reduces to
\begin{equation}\label{eq:u3}
  u_1(\OT)u_1(-\OT)\;=\;-ic_1\;{\pmb{\mathbb{M}}}_{12}(\OT)\;.
\end{equation}
This is impossible, because the left-hand side is even wrt $\OT$, while the right-hand side is odd for $\pmb{\mathbb{M}}_{12}(-\OT)=-\pmb{\mathbb{M}}_{12}(\OT)$.

(ii) $d_3=\cos(\theta_1)+\cos(\theta_2)$. We generalize the proof of case (i). Assume the spectrum to be pure point. Then one can easily check that for $\OT$ in the contour ${\cal C}$ defined as $\{\OT\in \mathbb{T}^2: \cos(\theta_1)+\cos(\theta_2)=0\}$,
Eq.~(\ref{eq:u3}) still follows (the constant $c_1$ is in general different). On the other hand, because the contour ${\cal C}$ has the inversion symmetry, if $\OT\in {\cal C}$ then $-\OT\in {\cal C}$. So, by the same arguments as above, Eq.~(\ref{eq:u3}) cannot be true for $\OT\in {\cal C}$. This contradiction proves the continuity  of the spectrum of
$\mathbb U$. $\Box$\\

\subsection{Proof of Corollary \ref{coro:modI}}
\label{sec:prctrex}

It is sufficient to consider the eigenfunction $\pmb{v}^+_{\boldsymbol{0}}$, which will be denoted by $\pmb{v}$ to make all formulae compact. From Eq.~(\ref{eq:dich2}) it follows that
\begin{equation}\label{eq:eigv}
  \pmb{M}_{\frac{\pi}{2}}(\OT)\,\pmb{v}(\OT-\pmb\omega)\;=\;ie^{i\pmb r\pmb\cdot\pmb\omega/2}\;\pmb{v}(\OT).
\end{equation}
Moreover, applying Eqs.~(\ref{eq:AB}) and (\ref{eq:topeig}) gives
\begin{equation}\label{eq:eigv1}
  \pmb{v}(\OT-\pmb\omega)\;=\;c^+_{\boldsymbol{0}}\;e^{-i\pmb r\pmb\cdot\OT}\;\pmb\sigma_3\;\pmb{v}(-\OT)\;.
\end{equation}
With the substitution of Eq.~(\ref{eq:eigv1}), Eq.~(\ref{eq:eigv}) reduces to
\begin{equation}
\label{eq:dtre0}
-i\pmb M_{\frac{\pi}{2}}(\OT) \pmb\sigma_3 \pmb{v}(-\OT)\;=\;(c^+_{\boldsymbol{0}})^{-1}e^{i\pmb r\pmb\cdot\pmb\omega/2}e^{i\pmb r\pmb\cdot\OT} \pmb{v}(\OT)\;.
\end{equation}
Let this equation be written at four points: $\pmb{\vartheta}_0\equiv(0,0)$, $\pmb \vartheta_1\equiv(0,\pi)$, $\pmb \vartheta_3\equiv(\pi,0)$, and $\pmb \vartheta_3\equiv(\pi,\pi)$. At each such point, Eq.~(\ref{eq:sm1}) yields that $i\pmb M_{\frac{\pi}{2}}(\pmb{\vartheta}_j)\pmb\sigma_3=d_3(\pmb{\vartheta}_j)\Id$. In addition,
\begin{equation}
\label{eq:dtre}
d_3(\pmb{\vartheta}_0)\;=\;-1\;,\;\;\;d_3(\pmb{\vartheta}_j)\;=\;1\;,\;\;(j=1,2,3)\;.
\end{equation}
So, if $\pmb{v}$ is continuous at $\pmb{\vartheta}_0$, then
$c^+_{\boldsymbol{0}}\,\pmb{v}(\pmb{\vartheta}_0)\;=\;e^{i\pmb r\pmb\cdot\pmb\omega/2}\pmb{v}(\pmb{\vartheta}_0 )$. This forces the choice
$c^+_{\boldsymbol{0}}=e^{i\pmb r\pmb\cdot\pmb\omega/2}$ [note that from Eq.~(\ref{eq:topeig}) $c^+_{\boldsymbol{0}}=\pm e^{i\pmb r\pmb\cdot\pmb\omega/2}$], because $\pmb{v}$ does not vanish at any point in $\mathbb{T}^2$. With this choice and the substitution of Eq.~(\ref{eq:dtre}), Eq.~(\ref{eq:dtre0}) entails:
\begin{equation}\label{eq:eigv2}
  \pmb{v}(\pmb{\vartheta}_j)\,=\,-e^{i\pmb r\pmb\cdot\pmb{\vartheta}_j}\, \pmb{v}(\pmb{\vartheta}_j),\;\;\;j=1,2,3,
\end{equation}
provided that $\pmb{v}$ is continuous at such points. If true for  $j=1,2$ , this equality requires both $r_1$ and $r_2$ [$\pmb r\equiv(r_1,r_2)$] to be  odd integers; but then it cannot be satisfied for $j=3$. This contradiction implies that the eigenfunction has a non-removable discontinuity at one at least of the four points $\pmb{\vartheta}_j$ ($j=0,1,2,3$). Using ergodicity of $\pmb\tau_{\pmb\omega}$ it is not difficult to prove that this forces the eigenfunction to be  actually discontinuous at all points. This will  not be shown here \cite{footnote_sing}.
Anyway, non-removable discontinuities cause delocalization of the eigenfunction, as they forbid
 finite expectation values for the momentum operators, which involve derivatives \cite{footnote_so}.

\begin{subsection}{Proof of Proposition \ref{propo:meanerg}}
\label{prop:ddelo}

By assumption, no eigenfunction belongs in the domain of the momentum operators, that is defined as $\{\ppsi\in {\cal H}: E_j(\ppsi)<+\infty\}$ where $j=1\, {\rm or}\, 2$ [recall Eq.~(\ref{eq:moment}) for the definition of $E_j(\ppsi)$]. Let the initial state $\ppsi$ be in this domain and $\ppsi_t=\U^t\,\ppsi$. For any $a>0$ we construct a subset of ${\cal H}$, defined as $K_a:=\{\ppsi\in {\cal H}: E_j(\ppsi)\leq a\}$. This subset is compact. Assume -- in
contradiction to Eq.~(\ref{eq:def_del}) -- that dynamical localization follows. Then $a$ can be chosen so that  $\ppsi_t\in K_a$, $\forall t \in \mathbb{Z}$. Let $e^{i\chi}$ be any eigenvalue of $\U$; obviously, $e^{-i\chi t}\ppsi_t\in K_a$. Now $E_j^{\frac{1}{2}}(\ppsi)$ is subaddittive, so we have:
\begin{gather}\label{eq:subadd}
  E_j^{\frac{1}{2}}(\tfrac{1}{t}\sum_{s=0}^{t-1}e^{-i\chi t}\ppsi_t)\leq
  \tfrac{1}{t}\sum_{s=0}^{t-1}E_j^{\tfrac{1}{2}}(e^{-i\chi t}\ppsi_t).
\end{gather}
Thus $E_j^{\frac{1}{2}}(\frac{1}{t}\sum_{s=0}^{t-1}e^{-i\chi t}\ppsi_t)\leq a$, and so
\begin{equation}\label{eq:addpsi}
  \tfrac{1}{t}\sum_{s=0}^{t-1}e^{-i\chi t}\ppsi_t\in K_a.
\end{equation}
Thanks to von Neumann's mean ergodic theorem \cite{Simon1}, in the limit $t\to +\infty$ the left-hand side of the above equation tends in ${\cal H}$ to the projection of the initial state $\ppsi$ on the $e^{-i\chi}$ eigenspace. So, since the spectrum is nondegenerate,
the limit of Eq.~(\ref{eq:addpsi}) is a scalar multiple of the eigenfunction. On the other hand, the limit
has to be in $K_a$ due to compactness; and this is a contradiction, because no eigenfunction of $\U$
can be in $K_a$. $\Box$\\
\end{subsection}

\section{Conclusion}
\label{sec:con}

In this paper, we have shown that, when the $2$D Maryland model is endowed with spin 1/2, rich dynamical localization/delocalization phenomena arise. In particular, in the family of SM Model I, dynamical localization--delocalization transitions are found to appear at half-integer $\hbar_e^{-1}$. Moreover, as a spinful QKR exhibiting $\hbar_e$-driven IQHE is deformed continuously into the considered SM model, its topological phases are deformed continuously into the localized phases of the latter, although we have not been able to identify the topological invariant for the latter systems. Because dynamical properties of SM models and spinful QKR are very different and so the topological theory \cite{Tian14,Tian16} developed for spinful QKR does not apply here, the present findings imply that the striking similarity between IQHE on the one side and spinful QKR on the other has a deeper origin, and may be carried to a broader class of dynamical spin systems.

We have uncovered a self-duality underlying the observed dynamical transition: the dynamically localized phases on both sides of a critical point are dual in accordance with Proposition \ref{propo:specsim}, while the critical delocalized phase is self-dual and as such bears an emergent unitary symmetry. This scenario of dynamical transition is conceptually different from Anderson-like dynamical transition observed in high-dimensional spinless QKR systems \cite{Casati89,Garreau08,Tian11}; it goes far beyond the Landau-Ginzburg paradigm of phase transition which finds its origin in symmetry breaking. It remains a prominent issue to explore the applications of this new transition scenario in other quantum dynamical systems and its relations to topological transitions in spinful QKR.

Finally, we note that the SM model has an equivalent classical dynamical system for any $\hbar_e$ values. This classical system belongs to a special class of skew product on $\TM^D\times G$ with $G$ a compact Lie group in general, which are currently investigated by mathematicians: SM models correspond to $D=2$ and $G=SU(2)$. In the language of skew-product systems, the transition in quantum dynamics is translated into the transition in the stability of classical trajectories in phase space. In-depth investigations of this aspect may allow one to view topological quantum phenomena such as IQHE from the perspectives of skew products on $\TM^D\times G$ and {\it vice versa}.

\section*{Acknowledgements}

In memory of Shmuel Fishman, who first proposed investigation of SM models.
Project Nos. 11535011, 11925507 and 11947302 supported by NSFC.\\

\appendix

\section{Spectral theory and cohomology}
\label{sec:cohomology}

Two maps $\boldsymbol{M},\boldsymbol{M}':{\mathbb T}^2\to SU(2)$ can be said to be $\OM$-{\it cohomologous} \cite{aar}, if there is a map $\boldsymbol{V}:{\mathbb T}^2\to SU(2)$ so that the classical dynamical systems on $\Omega$ corresponding to $\boldsymbol{M},\boldsymbol{M}'$ are isomorphic under the fiber-preserving isomorphism $(\OT, \boldsymbol{\phi})\mapsto(\OT, \boldsymbol{V}({\OT})\boldsymbol{\phi})$. Equivalently,
if, $\forall{\OT}$:
\begin{equation}\label{eq:coho}
  \boldsymbol{M}(\OT)=\boldsymbol{V}(\OT)\boldsymbol{M}'(\OT)\boldsymbol{V}^{-1}(\OT-\OM).
\end{equation}
In other words, $\boldsymbol{M}$, $\boldsymbol{M}'$ are $\OM$-cohomologous if, and only if,  the corresponding operators $\U_{M,\OM}$, $\U_{M',\OM}$ are unitarily equivalent: $\U_{M,\OM}=\boldsymbol{V}\U_{M',\OM}\boldsymbol{V}^{-1}$, where $\boldsymbol{V}$ is a unitary fiber-preserving operator and $\U_{M,\OM}:=\boldsymbol{M}\boldsymbol{T}_{\OM}$ with the Floquet operator $\U_{\alpha,\OM}=\boldsymbol{M}_\alpha\boldsymbol{T}_{\OM}$ as a special case.

It is then easy to see that $\U_{\alpha,\OM}$ has a pure point spectrum (\ref{eq:specq}) if, and only if, $\boldsymbol{M}_\alpha$ is $\OM$-cohomologous to the constant $SU(2)$ matrix that is diagonal on the canonical basis, with eigenvalues $e^{i\chi}, e^{-i\chi}$; where $\boldsymbol{V} (\OT)$ is the $SU(2)$ matrix that changes the local $(\pmb u(\OT), \boldsymbol{\rho}\,\pmb u(\OT))$ basis into the canonical basis. This leads to the following formulation of the quantum dynamics:
\begin{eqnarray}
\label{eq:cohq}
  (\U_{\alpha,\OM}\pmb\psi)(\OT) &=& e^{i\chi}[\pmb u(\OT-\OM)\pmb\cdot\pmb\psi(\OT-\OM)]\pmb u(\OT) \nonumber\\
  &+& e^{-i\chi}[\boldsymbol{\rho}\pmb u(\OT-\OM)\pmb\cdot\pmb\psi(\OT-\OM)]\boldsymbol{\rho}\pmb u(\OT).\quad
\end{eqnarray}

\section{Fishman-Grempel-Prange formulation of SM models}
\label{sec:App}

 Let ${\pmb U}^{\diamond}$ denote the inverse Cayley transform  of a unitary operator $\pmb U$ in $\cal H$: that is, the self-adjoint operator which is defined by
\begin{eqnarray}
{\pmb U}^{\diamond} \;=\;i\,(\mathbb{I}\,-\,\pmb U)\,(\mathbb{I}\,+\,\pmb U)^{-1}\;,\nonumber\\
 \pmb U\,=\; (i\;-\;{\pmb U}^{\diamond})\,(i\;+\;{\pmb U}^{\diamond})^{-1}\,.
\end{eqnarray}
The eigenvalue equation $\pmb M_{\alpha}\pmb T_{\pmb\omega}\pmb u\;=\;e^{i\lambda}\pmb u$ is easily seen to be equivalent to $\pmb H\pmb u=0$, where
$\pmb H={\pmb M}_{\alpha}^{\diamond} +(\pmb T_{\pmb\omega}e^{-i\lambda})^{\diamond}$. This formulation was first worked out by Fishman, Grempel and Prange for standard QKR \cite{Fishman82a}.
In the
momentum-spin representation, with basis vectors $|
\boldsymbol{N}, s\rangle$ ($
\boldsymbol{N}\in\ZM^2$, $s=1,2$),
matrix elements of $\pmb H$ are easily computed:
\begin{eqnarray}
\label{eq:hop}
\langle\boldsymbol{N},s|\pmb H|\boldsymbol{N}',s'\rangle\;=\;\delta_{ss'}\delta_{\boldsymbol{N}}\boldsymbol{N}'\tan \boldsymbol{N}\pmb\cdot\pmb\omega/2-\lambda/2)\;+\;\nonumber\\
-\;\tan (\alpha/2)\,\sum\limits_{k=1}^3\langle s|\pmb\sigma_k|s'\rangle\;\hat d_k(\boldsymbol{N}-\boldsymbol{N}')\,,\quad
\end{eqnarray}
where $\hat d_k(.)$ are the 2D Fourier coefficients of the function $d_k(\pmb\theta)$.  This can be read as a Hamiltonian for a spin $1/2$ particle on a $2$D discrete lattice, and resembles a tight-binding model in solid state physics. Like in the spinless Maryland model,  the first term is a spin-independent, on-site potential in the $\pmb N$-space. It is quasi-periodic, so long as $\pmb\omega$ is incommensurate. The second
term describes hopping as well as spin coupling between different sites. It   decays with the distance: $|\boldsymbol{N}-\boldsymbol{N}'|$ the faster,  the smoother the functions $d_k(\OT)$ are.



\begin{thebibliography}{}
\bibitem{footnote_SM} The Maryland models studied in the present work are a special class of quantum kicked systems. However this denotation is more frequently reserved to  certain tight-binding-like models in a quasi-periodic potential. Such models are related to the kicked systems by a construction \cite{Fishman82a}, that can be replicated for the SM models, as shown in Appendix \ref{sec:App}.
\bibitem{nature} The term ``duality" is here used in a sense which is similar, though not identical, to Ref.~\cite{Vitelli20}. Not so for the term ``symmetry".
\bibitem{footnote_x} The triple $(1,x,x^2)$ is the ``most incommensurate" triple of real numbers~\cite{Casati89}.
\bibitem{linstab} It is worth recalling that in classical integrable systems instability is generically linear in time.
\bibitem{footnote_ph} In this discussion we are excluding  pathological exceptions. These may be expected to occur when $\pmb\omega$ is close to commensurate.
\bibitem{footnote_bound} Though formally $1$D, the dynamics (\ref{eq:qhe1Dn1}) are not subject to this bound, because their one-period evolution operators explicitly depend on time $t$.
\bibitem{footnote_sing} It follows from minimality of the shift $\pmb\tau_{\omega}$, because the set of the points of discontinuity is closed and invariant.
\end{thebibliography}
\end{document}